\documentclass[lettersize,journal]{IEEEtran}
\usepackage{amsmath,amssymb}
\usepackage{multirow}
\usepackage{booktabs}
\usepackage[hidelinks]{hyperref}  
\usepackage{hyperref}
\usepackage{bm}
\usepackage{graphicx,graphics,color,psfrag}
\usepackage{cite,balance}
\usepackage{caption}
\allowdisplaybreaks
\usepackage{algorithm}
\usepackage{algorithmic}
\usepackage{accents}
\usepackage{amsthm}
\usepackage{url}
\usepackage[english]{babel}
\usepackage{multirow}
\usepackage{enumerate}
\usepackage{cases}
\usepackage{stfloats}
\usepackage{subfigure}
\usepackage{dsfont}
\usepackage{color,soul}
\usepackage{amsfonts}
\usepackage{cite,graphicx,amsmath,amssymb}
\usepackage{fancyhdr}
\usepackage{hhline}
\usepackage{graphicx,graphics}
\usepackage{array,color}
\usepackage{mathtools}
\usepackage{amsmath}

\graphicspath{ 
    {Results_figure/Train_result/Loss_vs_epochs/}
	{Results_figure/Train_result/Pattern/}  
	{Results_figure/Train_result/}  
	{figure/} 
}

\newtheorem{remark}{\bf \emph{\underline{Remark}}}
\newtheorem{example}{\bf \emph{\underline{Example}}}

\def\({\left(}
\def\){\right)}

\def\b0{{\mathbf{0}}}

\begin{document}
\captionsetup[figure]{name={Fig.},labelsep=period,singlelinecheck=off}  
	\title{\!Near-field\! Beam \!Training \!under\! Multi-path \!Channels:\! A Hybrid Learning-and-Optimization Approach}
	\author{Jiapeng~Li, 
		Changsheng~You, Guoliang~Cheng, Haobin~Sun,  Chao~Zhou, and Linglong~Dai\thanks{ 
			J. Li, C. You, G. Cheng, H. Sun, and C. Zhou are with the Department of Electronic and Electrical Engineering, Southern University of Science and Technology, Shenzhen 518055, China; G. Cheng is also with the Frontier Research Center, Peng Cheng Laboratory, Shenzhen 518055, China; 
			L. Dai is with the Department of Electronic Engineering and the State Key Laboratory of Space Network and Communications, Tsinghua University, Beijing 100084, China.
			 (e-mail: \{lijiapeng2023, chenggl2024, sunhb2023, zhouchao2024\}@mail.sustech.edu.cn; youcs@sustech.edu.cn; daill@tsinghua.edu.cn).
			(\emph{Corresponding author: Changsheng You.})}\vspace{-15pt}}
	\maketitle

\begin{abstract}

For extremely large-scale arrays (XL-arrays), the discrete Fourier transform (DFT) codebook, conventionally used in the far-field, has recently been employed for near-field beam training. However, most existing methods rely on the line-of-sight (LoS) dominant channel assumption, which may suffer degraded communication performance when applied to the general \emph{multi-path} scenario due to the more complex received signal power pattern at the user.
To address this issue, we propose in this paper a new \emph{hybrid} \emph{learning-and-optimization}-based beam training method that first leverages deep learning (DL) to obtain coarse channel parameter estimates, and then refines them via a model-based optimization algorithm, hence achieving high-accuracy estimation with low computational complexity.
Specifically, in the first stage, a tailored U-Net architecture is developed to learn the non-linear mapping from the received power pattern to coarse estimates of the angles and ranges of multi-path components.
In particular, the inherent \emph{permutation ambiguity} in multi-path parameter matching is effectively resolved by a permutation invariant training (PIT) strategy, while the unknown number of paths is estimated based on defined path existence logits.
In the second stage, we further propose an efficient particle swarm optimization method to refine the angular and range parameters within a confined search region; in the meanwhile, a Gerchberg-Saxton algorithm is used to retrieve multi-path channel gains from the received power pattern.
Last, numerical results demonstrate that the proposed hybrid design significantly outperforms various benchmarks in terms of parameter estimation accuracy and achievable rate, yet with low computational complexity.

	\end{abstract}

	\section{Introduction} \vspace{-2pt}
	\emph{Extremely large-scale arrays} (XL-arrays) have emerged as a promising technology to provide significantly high achievable rates to support emerging applications in the sixth-generation (6G) wireless networks, such as holographic communications, extended reality, and digital twins~\cite{10496996, 10663521}.
	With the dramatic increase in the number of antennas, users are more likely to fall into the near-field region, where radio propagation is characterized by spherical wavefronts rather than planar ones~\cite{10858129, 9913211}.
	This introduces a new degree-of-freedom in the range domain that unlocks the capability of location-aware \emph{beam-focusing}, allowing energy concentration~at specific locations instead of specific directions~\cite{10239282,10149471}.

	To achieve the benefits of beam-focusing, acquiring accurate channel state information (CSI) is indispensable, which is rather difficult in near-field systems as compared to far-field counterparts since both the angle and range parameters need to be estimated. Generally speaking, CSI acquisition methods can be categorized into two main paradigms, namely, explicit channel estimation~\cite{10149471,9693928} and implicit beam training~\cite{9913211,10500334,wang2025sparsityaware}. In this paper, we focus on beam training in near-field systems, which is particularly important in high-frequency bands for establishing a reliable link before efficient data transmissions based on predefined codebooks.

	\subsection{Related Works} \vspace{-2pt}
	Existing near-field beam training methods can be broadly categorized into two types, namely, the power-pattern-based and deep learning (DL)-based methods.
	\subsubsection{Power-pattern-based Near-field Beam Training}
	Power-pattern-based methods aim to infer key near-field channel parameters from structural features of the received power pattern at the user after beam sweeping, without explicitly reconstructing full channel matrix.
	Among others, a straightforward approach is a two-dimensional (2D) exhaustive search for both the angle and range in a polar-domain codebook~\cite{10858129}, which, however, incurs prohibitively high beam training overhead.

	To tackle this issue, a representative line of works exploited the structural regularity of the received power pattern under beam sweeping.
	For example, a two-phase near-field beam training method was proposed in~\cite{9913211}, which first estimates user angle through angle-domain beam sweeping based on a far-field oriented discrete Fourier transform (DFT) codebook and then estimates user range in a second stage with a polar-domain codebook.
	The required beam training overhead was further reduced in~\cite{10500334} by jointly estimating the angle and range based on a defined \emph{angular support} under the line-of-sight (LoS) path assumption, whose middle angle and width are used to infer the user angle and range, respectively.
	Furthermore, a low-complexity method was proposed in~\cite{wang2025low}, which efficiently obtains the near-field user range by analytically resolving the beam pattern generated by the DFT codebook.
	Besides the above methods using DFT codebook, other works reduced the beam training overhead by using different beam-design strategies.
	For instance, the authors in~\cite{10239282} proposed a coarse-to-fine procedure, which first generates wide beams and then gradually resolves finer-grained user angle and range parameters with narrow beams, thereby achieving low-overhead beam training.
	Likewise, the authors in~\cite{10365224} designed a multi-resolution near-field codebook, where wide beams and narrow beams are jointly designed to realize hierarchical search directly over the angle-range domain. 
	In addition, the multi-beam training method was proposed in~\cite{11235608, wang2025sparsityaware} to simultaneously form multiple focusing beams, which first identifies several candidate user locations by multi-beam sweeping and then determines the true user location with a few single-beam pilots, thereby  reducing the beam training~overhead.

	\subsubsection{DL-based Near-field Beam Training}
	In contrast to the above power-pattern-based methods, DL-based near-field beam training treats the inference from received signal powers to channel parameters as a data-driven learning problem. The main motivation is that the received power pattern is jointly determined by the highly nonlinear relationship among the angle, range, and channel gain parameters, for which an accurate and tractable analytical characterization is generally difficult to obtain in practical scenarios. By leveraging the strong function-approximation capability of deep neural networks, DL-based methods can effectively learn this intricate mapping from labeled beam-training data.
	Most existing works focused on LoS-only or LoS-dominant channels and formulated near-field beam training as a codeword classification problem, which can be efficiently solved via supervised learning.

	Specifically, the authors in~\cite{9903646} first proposed a DL-based near-field beam training method, where the received signal powers of far-field wide beams were exploited to separately estimate the angle and range indices of the optimal near-field codeword.
	This method was further improved in~\cite{10163797}, where a deep neural network (DNN) was trained over a near-field codebook with explicit angle-range information, such that the angle and range could be jointly estimated rather than inferred separately.
	Different from these codeword classification-based designs, the beamforming vector was directly learnt in~\cite{10682562}, where the achievable rate is set as the learning objective, thereby extending the codebook-based beam training to codebook-free beamformer generation.
	More recently, the authors in~\cite{wang2025tracking} further integrated beam training and beam tracking into a unified DL framework, where a two-stage convolutional neural network (CNN) was used for coarse-to-fine near-field beam training and a long-short-term-memory (LSTM) network was employed to exploit temporal correlation for mobility tracking.

	However, in view of the above works, the power-pattern-based and DL-based near-field beam training methods mostly have fundamental limitations when applied to general multi-path scenarios, which can be summarized as follows.
		\begin{itemize}
		\item \textbf{(Rely on the LoS-dominant channel assumption)}
		Existing near-field beam training methods rely heavily on the LoS or LoS-dominant channel assumption, which may not be valid in practical multi-path scenarios.
		In particular, if there exist strong non-line-of-sight (NLoS) paths, they may distort the received power pattern of the LoS channel case, exhibiting e.g., spurious power fluctuations and resulting in degraded beam training accuracy.
		\item \textbf{(Unknown number of dominant channel paths)}
		Most existing DL-based methods assumed a prior knowledge of the channel sparsity level (i.e., the number of paths). In practical scenarios, the number of scatterers is generally unknown and time-varying. A mismatch in the number~of paths may result in either over-fitting (estimating noise as paths) or under-fitting (missing dominant paths) issues, both of which degrade the estimation performance.
		\item \textbf{(Permutation ambiguity in estimated multi-path)}
		The physical multi-path channel is generally order-independent, whereas conventional neural networks typically generate order-sensitive sequences, where the estimation performance may be biased by the specific sorting of paths.
		This introduces permutation ambiguity~\cite{yu2017permutation}, where the indices of estimated paths may not well match with those of ground-truth paths.
		For instance, if path parameters are accurately estimated but misaligned in sequence, standard loss functions may incorrectly penalize this as a severe error. 
		Such permutation ambiguity also confuses optimizers, causing training instability and potential convergence~failure.
	\end{itemize}

	\vspace{-4pt}
	\subsection{Contributions} \vspace{-2pt}
	To address the above issues, we study efficient beam training design for near-field systems under the general multi-path channels. Specifically, we consider a practical scenario where the base station (BS) performs beam sweeping based on a DFT codebook. 
	While traditionally utilized in the far-field, the extension of DFT-based sweeping to the near-field has recently gained upsurging attention, as it facilitates a \emph{field-agnostic} beam training framework that reconciles the disparate characteristics of both propagation regimes.
	Under this setup, we aim to estimate key multi-path channel parameters from the received signal powers, including the angles, ranges, and complex-valued gains of the user and scatterers.
	Our main contributions are summarized as follows.
	\begin{itemize}

		\item
		First, we propose a new \emph{hybrid learning-and-optimization} method for near-field beam training under general multi-path channels, which first leverages DL to obtain coarse channel parameter estimates, and then refines them via a model-based optimization algorithm, hence achieving high-accuracy estimation with low computational complexity.
		Note that unlike existing methods tailored for LoS-only or LoS-dominant scenarios, the proposed framework accounts for the intricate nonlinear coupling and signal superposition among multiple paths in the received power pattern, which renders existing near-field beam training methods suboptimal or even inapplicable. 

		\item 
		Second, we propose efficient methods for the two stages. Specifically, for the first stage, we design a customized U-Net to learn the highly complex and nonlinear mapping from the received power pattern to coarse multi-path position estimates, where permutation invariant training (PIT) strategy is adopted to resolve the inherent permutation ambiguity in multi-path estimation and path existence logits are introduced to infer the unknown number of dominant paths.		
		Based on the statistical error characteristics of coarse estimates, we construct a confined search region for the key channel parameters. Subsequently, in the second stage, a customized particle swarm optimization (PSO) algorithm is developed to refine the angle-range parameters within this confined search region, while a Gerchberg-Saxton (GS) algorithm is employed to solve the corresponding phase retrieval problem and recover the complex-valued channel gains.	
		\item
		Last, extensive numerical results are provided to verify the effectiveness of the proposed method for near-field multi-path beam training. 
		It is shown that the proposed method can attain an achievable rate close to its upper bound based on perfect CSI under various simulation settings, yet it requires low computational complexity.
		Moreover, compared with the existing LoS-based methods, which suffer from severe model mismatch in multi-path scenarios, the proposed method achieves substantially improved estimation accuracy and achievable rate. 
	\end{itemize}

	\emph{Organization:} The remainder of this paper is organized as follows: Section~\ref{II} introduces the system model and problem formulation for near-field multi-path beam training. Section~\ref{III} presents the main limitations of two existing benchmark schemes and summarizes the proposed hybrid learning-and-optimization method.
	The proposed two stages are detailed in Sections~\ref{IV} and ~\ref{subsec:pso_refinement}, respectively.
	Numerical results are presented in Section~\ref{sec:results}, with conclusions provided in Section~\ref{sec:conclusion}.

	\section{System Model and Problem Formulation}	
	\label{II}
\begin{figure}[t]
		\centering
		\includegraphics[width=0.9\linewidth]{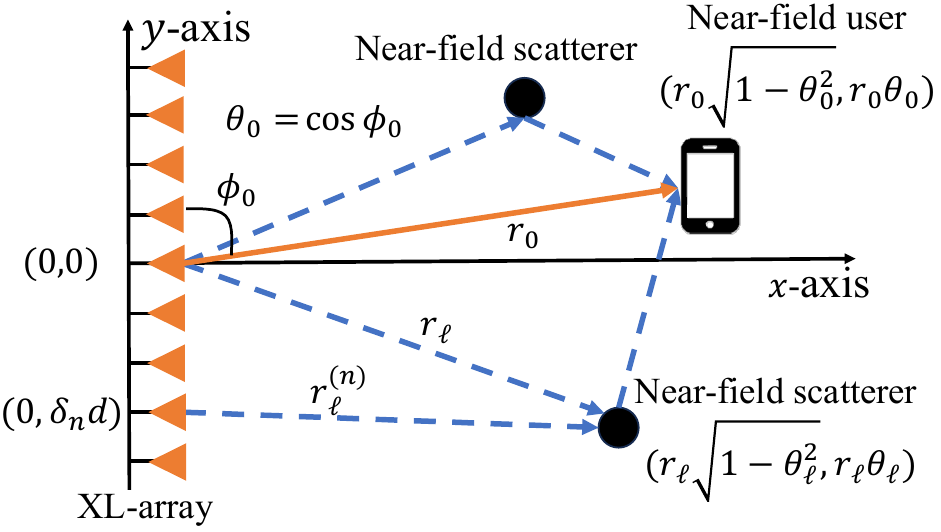}\vspace{-2pt}
		\caption{\centering A narrow-band XL-array communication system.}
		\label{MODEL}\vspace{-15pt}
	\end{figure}

	We consider downlink beam training for a narrow-band XL-array system as shown in Fig. \ref{MODEL}, where a BS equipped with an $N$-antenna uniform linear array (ULA) serves a single-antenna user.\footnote{The proposed beam training method can be directly applied to the multi-user case by estimating channels at individual users.} 
	Without loss of generality, the ULA is placed along the $y$-axis, where the $n$-th antenna is placed on $(0,\delta_nd)$, with $\delta_n=\frac{2n-N-1}{2}$, $n\in \mathcal{N} \triangleq\left\{1,2,\dots, N\right\}$ and $d=\frac{\lambda}{2}$  denoting the inter-antenna spacing.
	
	\vspace{-2pt}
	\subsection{Near-field Channel Model} \vspace{-2pt}
	The user is assumed to be located in the radiative near-field region of the BS.
	Based on the spherical wavefront propagation model, the near-field multi-path channel from the BS to the user can be modeled as~\cite{9913211}
	\begin{equation} \label{eq:total_channel_kappa} \vspace{-2pt}
		\mathbf{h}^{H}_{\rm {near}} =  g_{0}\mathbf{b}^{H} (\theta_0, r_0) 
		+  \sum_{\ell=1}^{L} g_{\ell} \mathbf{b}^{H}(\theta_{\ell}, r_{\ell}),
	\end{equation}
	which consists of one LoS path (represented by $\ell=0$) and $L$ NLoS paths (with $\ell \ge 1$).
	Herein, the parameters $g_{\ell}$, $\theta_{\ell} \triangleq \cos\phi_\ell$, and $r_{\ell}$ denote the complex-valued channel gain, spatial angle, and range of path $\ell$, respectively.
	In addition, $\mathbf{b}(\theta_{\ell},r_{\ell})\in \mathbb{C}^{N\times 1}, \ell \in \{0, 1,\dots,L\}$ denotes the near-field steering vector, modeled as~\cite{9913211, 10500334} 
	\begin{equation} \label{eq:steering_vector_H} 
		\mathbf{b}(\theta_{\ell}, r_{\ell}) = \Big[ e^{-\jmath \frac{2\pi}{\lambda}(r_{\ell}^{(1)}-r_{\ell})}, \ldots, e^{-\jmath\frac{2\pi}{\lambda}(r_{\ell}^{(N)}- r_{\ell})} \Big]^T, \end{equation} 
	where $r_{\ell}^{(n)} = \sqrt{r_\ell^2 +(\delta_nd)^2 - 2 r_\ell \theta_\ell \delta_nd}$ denotes the range between the $n$-th antenna and the user/scatterer.

	
	\subsection{Signal Model} \vspace{-2pt}
	We consider a practical two-phase transmission protocol. Specifically, in the first phase, the BS employs a predefined codebook for downlink beam training.
	Based on the received signal powers, the user estimates key channel parameters and then feeds them back to the BS.
	Then, in the second phase, the BS designs its transmit beamforming for downlink data transmissions based on estimated CSI.
	
	\subsubsection{\underline{\textbf{Phase 1: Downlink beam training}}}

	For downlink beam training, we consider the DFT codebook, which is a well-known codebook in far-field beam training due to its orthogonality and low implementation complexity.
	Recently, the DFT codebook was also shown to be effective for near-field beam training to jointly estimate the angle and range parameters~\cite{9913211, 10500334,wang2025low}, which requires much lower beam training overhead than the polar-domain codebook.
	As such, the DFT codebook can be regarded as a \emph{universal} codebook applicable to both near-field and far-field beam training. 
	
	Let $\mathbf{V}_{\rm DFT}=[\mathbf{v}_{1},\mathbf{v}_{2},\ldots,\mathbf{v}_{N}] \in \mathbb{C}^{N\times N}$ denote the DFT codebook used for beam training, which consists of $N$ orthogonal beamforming vectors that uniformly sample the spatial angle domain. The $n$-th vector is given by
	\begin{align}
		\!\!\!\!\mathbf{v}_{n} \!=\! \mathbf{a}(\varphi_{n}) \!\triangleq \! \frac{1}{\sqrt{N}}
		\Big[1,e^{-\jmath\pi\varphi_{n}},\ldots,e^{-\jmath\pi(N\!-\!1)\varphi_{n}}
		\Big]^T\!\!,\forall n\!\in\! \mathcal{N}\!,\!
	\end{align}
	where $\varphi_{n} = \frac{2n-N-1}{N},n\in\mathcal{N}$ denotes the sampled spatial angle of the $n$-th beam codeword. 
	Let $s\in \mathbb{C}$ denote the transmitted pilot signal with transmit power $P_{\rm t}$. 
	Then, the received signal at the user given the BS beamforming vector $\mathbf{v}_n$, can be expressed as 
	\begin{align} \label{eq:y_training}
		y(\mathbf{v}_{n}) = \mathbf{h}_{\rm near}^{H}\mathbf{v}_{n}s + z, \forall n\in \mathcal{N},
	\end{align}
	where $z\sim\mathcal{CN}(0,\sigma^2)$ is the received additive white Gaussian noise (AWGN) at the user with $\sigma^2$ denoting the noise power.
	As such, the received signal power under the beam sweeping of DFT codebook is given by
	\begin{align}\label{power_vector}
    p_n=|\mathbf{h}_{\rm near}^{H} \mathbf{v}_{n}s + z|^2, \forall n \in \mathcal{N}.
	\end{align} 
	Based on the received power pattern $\mathbf{p}_{\rm DFT} \triangleq [p_1, \dots, p_N]^T \in \mathbb{R}^{N\times 1}$, key multi-path channel parameters (or the equivalent channel) are estimated, including the angles and ranges of the user and scatterers $\bar{\boldsymbol{\theta}} = [\bar{\theta}_0,\dots,\bar{\theta}_L]^T$ and $\bar{\mathbf{r}} = [\bar{r}_0,\dots,\bar{r}_L]^T$, respectively, as well as the corresponding complex-valued channel gains $\bar{\mathbf{g}} = [\bar{g}_0,\dots,\bar{g}_L]^T$.
	For brevity, the overall estimated  parameters are represented as $\bar{\boldsymbol{\eta}}=[\bar{\boldsymbol{\theta}}^T, \bar{\mathbf{r}}^T]^T$ and~$\bar{\mathbf{g}}$.

	\subsubsection{\underline{\textbf{Phase 2: Data transmission}}}
	For Phase 2, given the estimated $\bar{\boldsymbol{\eta}}$ and $\bar{\mathbf{g}}$, the BS-user channel can be constructed
	\begin{align}
		\!\!\bar{\mathbf{h}}_{\rm near}^{H}(\bar{\boldsymbol{\eta}},\bar{\mathbf{g}})
		\!=\!
		\bar{g}_0\,\mathbf{b}^{H}(\bar{\theta}_0,\bar{r}_0)
		+
		\sum_{\ell=1}^{L}
		\bar{g}_\ell\,\mathbf{b}^{H}(\bar{\theta}_\ell,\bar{r}_\ell).\!\!
	\end{align}
	Based on $\bar{\mathbf{h}}_{\rm near}(\bar{\boldsymbol{\eta}},\bar{\mathbf{g}})$, the optimal beamforming vector for data transmission, denoted by $\mathbf{w}(\bar{\boldsymbol{\eta}},\bar{\mathbf{g}})$, can be designed based on the maximum-ratio transmission (MRT), i.e., $\mathbf{w}(\bar{\boldsymbol{\eta}},\bar{\mathbf{g}}) = \frac{\bar{\mathbf{h}}_{\rm near}(\bar{\boldsymbol{\eta}},\bar{\mathbf{g}})}{\| \bar{\mathbf{h}}_{\rm near}(\bar{\boldsymbol{\eta}},\bar{\mathbf{g}})\|_{2}}$, for which the corresponding achievable rate at the user in bits/second/Hertz (bps/Hz) is given by
	\begin{align}
		R = \log_2 \Big(1 + {P_{\rm t} |\mathbf{h}_{\rm near}^H \mathbf{w}(\bar{\boldsymbol{\eta}},\bar{\mathbf{g}})|^2}/{\sigma^2}\Big).
	\end{align}

	\subsection{Problem Formulation}\label{II_C} \vspace{-2pt}
	The aim of near-field multi-path beam training is to estimate key multi-path channel parameters based on the received power pattern, which is used to design the BS transmit beamforming for maximizing the user achievable rate.\footnote{ 
	The beamforming strategy adopted here shares a similar framework with the 5G NR Type II codebook mechanism~\cite{11346973}. 
	Both approaches achieve linear beam combination through explicit amplitude and phase modulation, thereby providing high-precision feedback of multi-path parameters for effective beamforming.}
	Specifically, according to \eqref{eq:total_channel_kappa} and \eqref{power_vector}, there exists an intrinsic mapping between $\{\boldsymbol{\eta}, \mathbf{g}\}$ and $\mathbf{p}_{\rm DFT}$. 
	Mathematically, the near-field multi-path beam training problem can be expressed as	
	\begin{align} \label{f_inv}
		\{\bar{\boldsymbol{\eta}}, \bar{\mathbf{g}}\} \Leftarrow  \mathcal{F}(\mathbf{p}_{\rm DFT}),
	\end{align}
	where $\mathcal{F}(\cdot)$ denotes the mapping function to be designed from the received power pattern to the channel parameters.
	
	Note that for the near-field multi-path channel case, it is generally challenging to obtain a specific mapping function $\mathcal{F}(\cdot)$ in closed form, due to the highly \emph{non-linear} and complex relationship between the received power pattern and multi-path channel parameters.

	\section{Benchmark Schemes and Proposed Framework} \vspace{-2pt} \label{III}
	In this section, we introduce two types of existing benchmark schemes for near-field beam training and point out their main limitations in the considered multi-path scenario.
	Besides, in order to overcome these limitations, we propose a new hybrid learning-and-optimization method.

	\vspace{-4pt}
	\subsection{Power-pattern-based Methods} \vspace{-2pt} \label{III_A}
	Power-pattern-based methods aim to estimate key channel parameters by analyzing \emph{structural} features (e.g., angular support) of the received power pattern, which was recently exploited for near-field single-path scenarios (see, e.g.,~\cite{9913211,10500334,wang2025low}).
	However, the estimation accuracy of these methods may be severely degraded in scenarios with strong-power multi-paths due to the following two reasons. 
	First, the angular supports of different paths may overlap in the received power pattern, which can be \emph{coherently added} or \emph{destructively counteracted}, depending on the phases of individual paths.
	Second, the random complex-valued path gains associated with NLoS components further perturb the received power pattern, posing a more challenging angle and range estimation problem.
	Thereby, the power-pattern-based methods are generally \emph{ineffective} in the near-field multi-path channels.

\begin{figure}[t]\centering
	\includegraphics[width=0.48\textwidth]{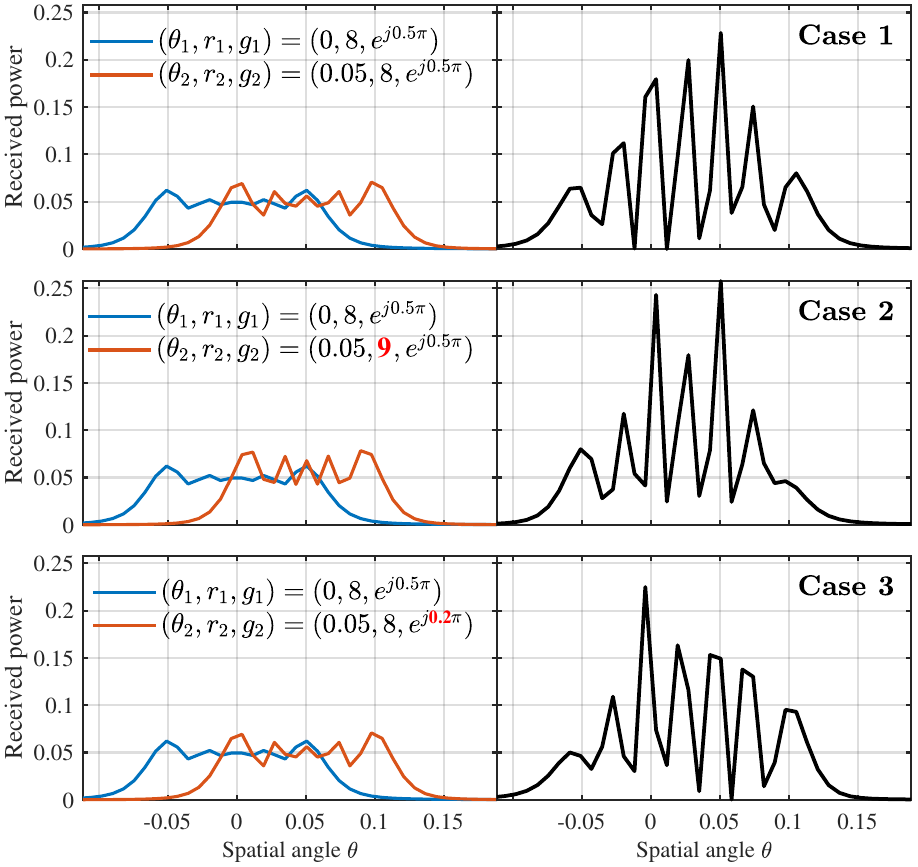}\vspace{-2pt}		
	\caption{Power pattern comparison.
	The left column shows the separated components defined by $\mathbf{h}^H_i = g_i \mathbf{b}^H(\theta_i,r_i)$ for $i=1,2$, while the right column shows the superposed channel $\mathbf{h}^H = \mathbf{h}^H_1 + \mathbf{h}^H_2$. Red text highlights parameter differences compared to Case 1, demonstrating the high sensitivity of signal superposition to slight parameter variations.
	}\vspace{-8pt}
	\label{fig:coherent_superposition}
	\end{figure}

	\begin{example}[Signal superposition issue]\label{prop:pattern} \emph{
		For illustration, we compare in Fig. \ref{fig:coherent_superposition} the received power pattern of two individual paths and their superposed ones under three different cases.
		Specifically, for Case 1 and Case 2, their spatial angles and complex-valued channel gains of the two paths are identical, while the ranges of the second paths differ slightly ($r_2 = 8$ meter (m) in Case 1 versus $r_2 = 9$~m in Case 2). 
		It is observed that the individual power patterns of the cases are similar, but their corresponding power patterns of superposed signals are quite different, since a small range variation can lead to significantly different near-field channel steering vectors, resulting in distinct received power patterns.
		Next, we compare Case 1 and Case 3, where the physical positions (both angle and range) of the two paths are identical, while the phases of their channel gains are different, i.e., $g_2 = e^{\jmath 0.5\pi}$ in Case 1 versus $g_2= e^{\jmath 0.2\pi}$ in Case 3. 
		One can observe that even with such a mild phase variation, the power patterns of their superposed signals differ significantly, due to the effect of phase in the multi-path signal superposition (i.e., in-phase superposition or out-of-phase cancellation). 
		Consequently, relying on theoretical analysis of power pattern to estimate channel parameters is highly challenging (if possible) in multi-path scenarios, as slightly different channel parameters can yield significantly distinct received power patterns.
	}
	\end{example}

	\vspace{-4pt}
	\subsection{DL-based Methods} \vspace{-2pt}\label{III_B}
	Alternatively, DL-based methods have also been proposed for near-field beam training~\cite{9903646,10163797,10682562}, which leverage powerful non-linear mapping capabilities of DL to infer channel parameters from the received powers.	
	Although DL-based methods can achieve fast inference in near-field beam training under the LoS channels~\cite{9903646, 10682562}, they may \emph{not} be able to provide accurate beam training in practical near-field multi-path channel scenarios due to the following reasons.
	\begin{itemize}
		\item \textbf{Unknown number of paths:} First, the existing DL-based methods for multi-path beam training have mostly assumed that the number of paths is known \emph{a priori},  which, however, may not be available in practice.
		\item \textbf{Permutation ambiguity:} 
		Second, conventional DL methods typically require a  deterministic 	matching relationship between the output channel paths and the ground-truth channel paths during model training.
		However, due to the unordered nature of multi-path channels, there is no predefined matching relationship, resulting in a \emph{permutation ambiguity} issue in estimating multi-path parameters.

		\item \textbf{Accuracy limitation:} Third, the existing DL-based methods usually employ the classification framework to estimate the best beam codeword based on a predefined near-field codebook~\cite{9903646}, which inevitably introduces quantization errors in estimating continuous-valued channel parameters, thus resulting in limited estimation accuracy.
	\end{itemize}

	\subsection{Proposed Hybrid Framework} \vspace{-2pt} \label{III_C}

	\begin{figure*}[t]
		\centering
		\includegraphics[width=0.9\linewidth]{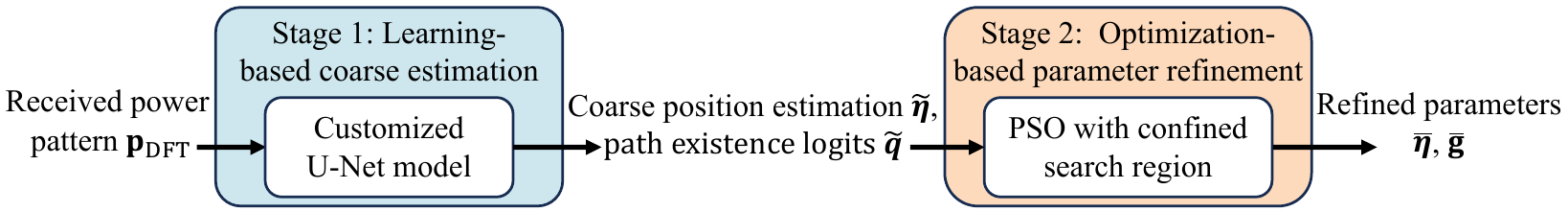} \vspace{-4pt}
		\caption{\centering The framework of proposed hybrid learning-and-optimization method.} \vspace{-10pt}
		\label{fig:algo_graph} 
	\end{figure*}
	To address the above issues, we propose in this paper a new \emph{hybrid learning-and-optimization} framework for near-field multi-path beam training, which synergistically fuses the efficiency of data-driven DL and the high precision of model-based optimization, thereby achieving accurate estimation of key channel parameters with low complexity.
	Specifically, as illustrated in Fig.~\ref{fig:algo_graph}, our key idea is to first estimate \emph{coarse} multi-path channel parameters by using DL techniques and then \emph{refine} the parameters by using optimization techniques.
	The main procedures are summarized below.
	\begin{itemize}
		\item \textbf{Stage 1 (Learning-based coarse estimation):} 
		For this stage, we design an efficient U-Net model to enable \emph{coarse} channel parameter estimation by using a PIT strategy to address the permutation ambiguity issue, estimating path existence logits to tackle the issue of an unknown number of paths, and employing a customized regression loss to reduce codebook quantization errors.
		Specifically, the model takes the obtained DFT power pattern $\mathbf{p}_{\rm DFT}$ as input, and outputs a coarse estimation of the multi-path angles and ranges $\tilde{\boldsymbol{\theta}}, \tilde{\mathbf{r}}$ (i.e., $\tilde{\boldsymbol{\eta}}$), as well as the corresponding path existence logit $\tilde{\mathbf{q}} = [\tilde{q}_0, \tilde{q}_1, \dots, \tilde{q}_L]^T$, which indicates the existence probability of each path.
		Moreover, the statistical characteristics of estimation errors are also obtained during model validation, based on which a \emph{confined} parameter search region $\mathcal{B}$ around the coarse estimate $\tilde{\boldsymbol{\eta}}$ is constructed for subsequent optimization.
		
		\item \textbf{Stage 2 (Optimization-based parameter refinement):} Given the coarse estimation $\tilde{\boldsymbol{\eta}}$ and confined search region $\mathcal{B}$, an optimization-based method is employed to obtain a high-precision estimation for the multi-path position parameters, denoted by $\bar{\boldsymbol{\eta}}$, by minimizing a defined discrepancy between the reconstructed and received power pattern based on the maximum likelihood principle.
		This problem is efficiently solved by a customized PSO algorithm, which effectively avoids poor local minima in the confined search region for the non-convex problem, achieving a high-quality solution with low complexity.
	\end{itemize}

	The details of our proposed hybrid method are presented in the next two sections.

	\section{Stage 1 for Learning-based Coarse Estimation}	\vspace{-2pt}\label{IV}
	
	This stage aims to leverage the powerful feature extraction capabilities of DL to  estimate coarse multi-path position parameters from the received power pattern $\mathbf{p}_{\rm DFT}$.
	It serves as a crucial preliminary step that significantly narrows down the search space for the subsequent optimization-based refinement.

	To achieve high-quality feature extraction, we employ a customized U-Net architecture as the backbone of our neural network model.
	Note that the U-Net architecture was originally proposed for biomedical image segmentation tasks~\cite{Ronneberger2015_UNet}.
	Its encoder–decoder structure with skip connections allows for effective multi-scale feature learning, making it applicable to various tasks, such as channel estimation and signal separation~\cite{9697063, 10945552}.
	For the considered near-field multi-path beam training problem, the U-Net architecture is well-suited due to the following reasons.
	First, the encoder in U-Net allows for progressively extracting features at different scales, enabling the neural network to capture both coarse dominant information (e.g., power peak of the LoS path) and detailed spatial patterns (e.g., power of superposed signals among multi-paths).
	Next, the skip connection structure is employed to concatenate features from corresponding layers in the encoder and decoder, so that the features extracted from the encoder layers can be directly transferred to the corresponding decoder layers. 
	Compared with traditional CNNs, the U-Net architecture can better preserve spatial information during feature extraction by fusing features from different scales, hence achieving more accurate channel parameter estimation than CNNs (to be numerically shown in Fig.~\ref{fig:error_vsSNR}).
	Moreover, unlike existing DL-based methods based on prior information of the number of channel paths, our proposed model does not require this prior knowledge, since it estimates the path existence probability for each potential path.
	In the following, we introduce the detailed model training method.


	\vspace{-6pt}
	\subsection{Data Preprocessing} \vspace{-2pt}
	To train the proposed U-Net model, we first need to generate a large dataset of labeled samples.
	Specifically, we generate a dataset consisting of $D$ samples, where each sample is generated by simulating near-field multi-path channels based on the system model described in Section~\ref{II}.
	For each sample, we randomly determine the number of NLoS paths $L$ within a maximum limit $L_{\max}$; the user and scatterer positions are randomly generated as $\{\theta_\ell, r_\ell\}_{\ell=0}^{L}$ within near-field region; and then compute the corresponding received power pattern $\mathbf{p}_{\rm DFT}$.
	Besides, to eliminate the scale disparity between the spatial angle (unitless) and range (in meters), the angle and range parameters are standardized to zero mean and unit variance before being fed into the neural network for training.
	During the online inference phase, the estimated standardized angle and range are transformed back to the spatial angle and physical range values for the subsequent optimization stage.
	This ensures balanced contributions from both angle and range parameters to the loss function, preventing bias caused by magnitude differences.
	Moreover, to handle the varying number of paths across different samples, we adopt a padding strategy where the angle and range parameters are padded with zeros up to the maximum number of paths~$L_{\max}$.
	Consequently, the ground-truth labels are constructed as $\{\theta_{\ell}, {r}_{\ell}, c_\ell\}_{\ell=0}^{L_{\max}}$, where $c_\ell$ denotes the path existence indicator with $c_\ell = 1$ indicating the existence of path $\ell$ and $c_\ell = 0$~otherwise.

	\vspace{-4pt}
	\subsection{Model Architecture}	\vspace{-2pt}
	\begin{figure*}[t]
		\centering
		\includegraphics[width=0.99\linewidth]{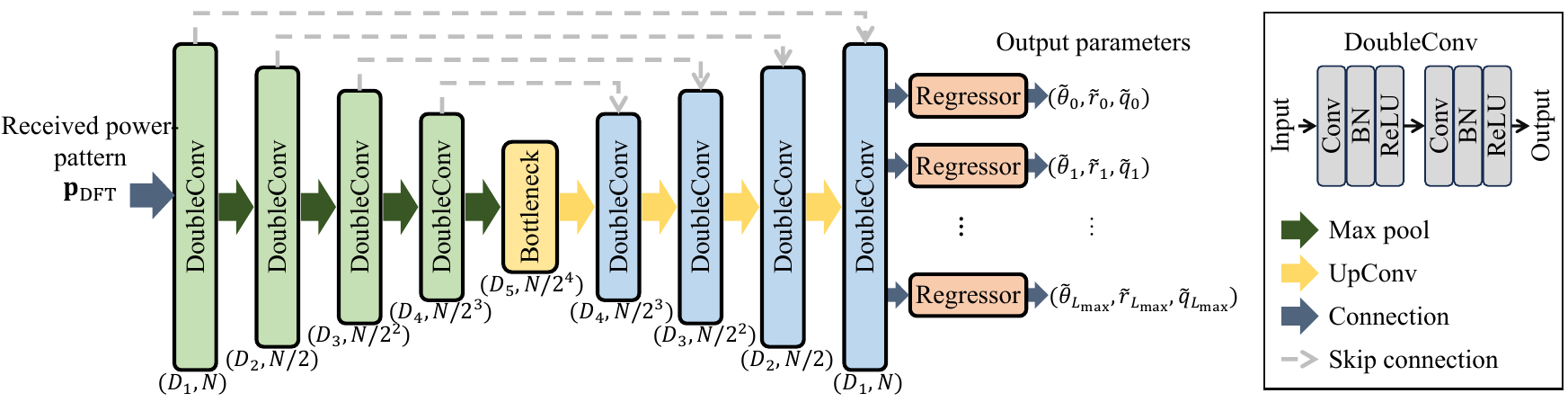} \vspace{-4pt}
		\caption{\centering U-Net architecture and training methodology for coarse estimation.}
		\label{fig:neural_model}
		\vspace{-16pt}
	\end{figure*}
	We adopt a 1D U-Net-based model as the core backbone to learn the mapping between the received power pattern $\mathbf{p}_{\rm DFT}$ (as input) and key channel parameters (as output) including the polar coordinates of user and scatterers $\tilde{\boldsymbol{\theta}}$, $\tilde{\mathbf{r}}$, as well as the path existence logits  $\tilde{\mathbf{q}} = [\tilde{q}_0,\dots,\tilde{q}_L]^T$.
	The overall architecture of the proposed U-Net model is illustrated in Fig.~\ref{fig:neural_model}.

	Specifically, the designed model comprises an encoder for feature extraction, a bottleneck for latent representation, and a decoder for parameter reconstruction. 
	The encoder stacks DoubleConv modules to progressively downsample the input received power pattern into higher-level semantic features along the spatial dimension $N$, condensing multi-path information into a global latent representation within the bottleneck. 
	The symmetric decoder employs transposed convolutions for upsampling, leveraging skip connections to fuse multi-scale features from the encoder, thereby preserving high-resolution spatial information. 
	Finally, a fully connected layer maps the restored features to the polar coordinates of paths and their existence logits.
	The fundamental DoubleConv block consists of two sequential operations, each including 1D convolution (Conv), batch normalization (BN), and rectified linear unit (ReLU) activation.
	Note that the U-Net model is employed solely to estimate angle and range parameters, while complex-valued channel gains will be reconstructed based on the estimated angle–range parameters (see Section~\ref{subsec:pso_refinement}).

\vspace{-4pt}
	\subsection{U-Net Training Method}\vspace{-2pt}
	For the proposed U-Net training method, we first judiciously design a loss function to address three key issues in near-field multi-path beam training, namely, the unknown number of paths, permutation ambiguity, and accuracy limitation.

	To this end, the designed loss function is composed of three losses, i.e., a regression loss for the position of LoS path, a permutation-invariant regression loss for the positions of NLoS paths, and a binary cross-entropy loss for path existence classification.
	These loss components are detailed as follows.

	\begin{itemize}
		\item \textbf{Regression loss for the LoS path:}
		In practical scenarios, the LoS path is easy to identify, as it has the strongest received power among all paths. 
		The regression loss for the LoS path refers to the Euclidean distance between the estimated and ground-truth user positions, given by
		\begin{equation} 
			\label{eq:LoS_loss_fun}
			\mathcal{L}_{\rm LoS} = \sqrt{(\tilde{\theta}_0-\theta_0)^2 + (\tilde{r}_0-{r}_0)^2}.
		\end{equation}

		\item \textbf{Permutation-invariant regression loss for NLoS paths:}
		Note that the received power pattern $\mathbf{p}_{\rm DFT}$ contains a LoS path and multiple NLoS paths, where the order of NLoS scatterers is generally unknown.
		This implies that even if the estimated angle and range parameters are accurate, the correspondence between estimated and ground-truth scatterer positions is not unique, hence resulting in the \emph{permutation ambiguity} issue in training the neural network.
		To tackle this issue, we propose a PIT-based method to design a permutation-invariant regression loss for the NLoS scatterers~\cite{yu2017permutation}, which emulates all possible label permutations and selects one permutation that minimizes the training loss in \eqref{eq:NLoS_loss_fun}.
		Specifically, to achieve this goal, we need to consider all possible permutations of the positions of ground-truth scatterers.
		As such, we construct a cost matrix $\mathbf{C}\in\mathbb{R}^{L_{\max}\times L_{\max}}$, where each element represents a squared error between the estimated and ground-truth  positions, i.e., $\tilde{\boldsymbol{\theta}}, \tilde{\mathbf{r}}$ and $\boldsymbol{\theta}, \mathbf{r}$, given by
			\begin{align}
				C_{\ell,m} = (\tilde{\theta}_\ell-\theta_{m})^2 + (\tilde{r}_\ell-{r}_{m})^2, \forall \ell,m \in \mathcal{L}_{\max},
			\end{align}
			where $\mathcal{L}_{\max} \triangleq \{1,2,\dots,L_{\max}\}$. 
			Let $x_{\ell,m}\in \{0,1\}$ define whether the estimated scatterer $\ell$ is matched with the ground-truth scatterer $m$ or not.
			Thus, the problem of finding the optimal permutation for achieving the minimum matching cost can be formulated as an assignment problem as follows
			\begin{subequations}\vspace{-2pt}
				\begin{align}
				\textbf{(P3)}: \min_{\{x_{\ell,m} \}} &\quad\sum_{\ell=1}^{L_{\max}} \sum_{m=1}^{L_{\max}} x_{\ell,m}C_{\ell,m} \nonumber\\
				\text{s.t.} &\quad \sum_{\ell=1}^{L_{\max}} x_{\ell,m} = 1, \forall m \in \mathcal{L}_{\max},\label{P1_C1}\\ 
				&\quad \sum_{m=1}^{L_{\max}} x_{\ell,m} = 1, \forall \ell \in \mathcal{L}_{\max}, \label{P1_C2}\\ 
				&\quad x_{\ell,m} \in \{0,1\}, \forall \ell,m \in \mathcal{L}_{\max}, \label{P1_C3}
				\end{align}
			\end{subequations}
			where constraints \eqref{P1_C1} and \eqref{P1_C2} ensure that each estimated scatterer is matched with exactly one ground-truth scatterer and vice versa.
			Problem (\textbf{P3}) is a classic linear sum assignment problem in combinatorial optimization, which can be solved in polynomial time by using the Hungarian algorithm, without the need to evaluate all permutations explicitly~\cite{yu2017permutation}.
			For the sake of brevity, the specific details of the Hungarian algorithm are omitted.
			We denote the optimal solution to Problem (\textbf{P3}) by $\{x_{\ell,m}^*\}$.
			As such, the optimal permutation function can be constructed based on the optimal assignment as follows
			\begin{align}\label{optimal_pi}	
				\pi(\ell) = m, \text{ if } x_{\ell,m}^* = 1.
			\end{align}
			Thus, we can obtain the permutation-invariant regression loss for NLoS paths based on the optimal permutation $\pi$ in \eqref{optimal_pi}.
		Mathematically, the \emph{permutation-invariant} regression loss for the positions of NLoS paths is defined as the Euclidean distance between the estimated and optimally matched ground-truth path positions, given by		
	\begin{equation} 
		\label{eq:NLoS_loss_fun}
		\!\!\mathcal{L}_{\rm Reg} \!=\! \sum_{\ell=1}^{L_{\max}} \sqrt{(\tilde{\theta}_\ell-\theta_{\pi(\ell) })^2 + (\tilde{r}_\ell-{r}_{\pi(\ell)})^2}.
	\end{equation}

		\item \textbf{Binary cross-entropy loss for path existence classification:}
		In order to estimate the number of dominant NLoS paths, we introduce a binary cross-entropy loss for each detected NLoS path.\footnote{Due to the unknown number of dominant NLoS paths, position regression losses in~\eqref{eq:LoS_loss_fun} and ~\eqref{eq:NLoS_loss_fun} are insufficient to identify valid channel components. 
		The path existence loss is essential to explicitly estimate the presence probability of each path, thereby enabling the accurate determination of the number of dominant multi-path components.}
		The loss is defined to measure the discrepancy between the estimated path existence logits $\tilde{\mathbf{q}}$ and the ground-truth path existence indicators $\mathbf{c} = [c_0, c_1, \dots, c_{L_{\max}}]^T$, given by
		\begin{align} \label{Loss_Cls}
			\!\!\!\!\! \mathcal{L}_{\rm Cls}\! = \!- \!\!\!\sum_{\ell=1}^{L_{\max}} \!\!\big[ c_{\pi(\ell)} \log\big(\sigma(\tilde{q}_\ell)\big) \!+ \!(1\!-\!c_{\pi(\ell)}) \log\big(1\!-\!\sigma(\tilde{q}_\ell)\big) \big]\!,\!\!
		\end{align}
		where $\sigma(x) = 1/(1 + e^{-x})$ denotes the sigmoid activation function that maps the logit $\tilde{q}_\ell$ by the neural network to a probability value between 0 and 1.
		Minimizing this loss encourages the U-Net model to output high-valued logits for present paths and low-valued logits for absent~ones.

	\end{itemize}

	Based on the above, the objective of the model training is to minimize the total loss function, which is defined as the sum of the LoS regression loss, the permutation-invariant NLoS regression loss, and the classification loss, given by
	\begin{align}\label{eq:total_loss_fun}
		\mathcal{L}_{\rm total} = \alpha_1\mathcal{L}_{\rm LoS} + \alpha_2\mathcal{L}_{\rm Reg} + \alpha_3\mathcal{L}_{\rm Cls},
	\end{align}
	where $\alpha_1$, $\alpha_2$, and $\alpha_3$ are weighting coefficients that balance the contributions of each loss component. The effectiveness of this designed loss function will be shown in Section~\ref{sec:results}.
	
	\vspace{-6pt}
	\subsection{Online Inference Procedure}\vspace{-2pt}
	After training, the U-Net model can be employed to infer the coarse multi-path position parameters based on the received power pattern $\mathbf{p}_{\rm DFT}$.
	Specifically, given the input $\mathbf{p}_{\rm DFT}$, the model outputs the coarse estimation of multi-path position parameters, i.e., the angles and ranges $\tilde{\boldsymbol{\theta}}$ and $\tilde{\mathbf{r}}$, as well as the path existence logits $\tilde{\mathbf{q}}$.
	To determine the number of dominant NLoS paths $\tilde{L}$, we apply a thresholding operation on the estimated path existence logits as follows
	\begin{equation}
		\tilde{L} = \sum_{\ell=1}^{L_{\max}} \mathbb{I}(\sigma(\tilde{q}_\ell) \geq \lambda_{\rm th}),
	\end{equation}
	where $\mathbb{I}(\cdot)$ is the indicator function that returns 1 if the condition is true and 0 otherwise, and $\lambda_{\rm th}$ is a predefined threshold.
	Let $k_1, k_2, \dots, k_{\tilde{L}}$ denote the indices of the detected $\tilde{L}$ paths, i.e., $\!\sigma(\tilde{q}_{k_i})\!\geq\! \lambda_{\rm th}, k_i \!\in\! \tilde{\mathcal{L}}\! \triangleq\! \{1,\dots,\!\tilde{L}\}$.
	Thus, the coarse angle-and-range estimates of the detected paths are given by $\tilde{\boldsymbol{\theta}}\! =\! [\tilde{\theta}_{k_1}, \dots, \tilde{\theta}_{k_{\tilde{L}}}]^T$ and $\tilde{\mathbf{r}} \!= \![\tilde{r}_{k_1}, \dots, \tilde{r}_{k_{\tilde{L}}}]^T$, respectively.
	
	\vspace{-2pt}
	\subsection{3$\sigma$-criterion Boundary}\vspace{-2pt}
	Besides generating the per-path coarse position parameter estimation ${\tilde{\boldsymbol{\eta}}} = [\tilde{\boldsymbol{\theta}}^T, \tilde{\mathbf{r}}^T]^T$, we can also obtain the statistical characteristics of the estimation error based on the trained model.
	Specifically, the expectation and variance of the estimation error for each path can be obtained during the U-Net model validation~\cite{10827219}, which are given by
	\begin{equation}
		\begin{aligned}
			\!\!\mu^{r}_{\ell} \!\!&= \mathbb{E}(\tilde{r}_\ell - {r}_\ell), \!\!\!\! &
			\sigma^{r}_{\ell} \!\!&= \sqrt{\mathbb{V}(\tilde{r}_\ell - {r}_\ell)}, \ell \in \tilde{\mathcal{L}},\\
			\!\!\mu^{\theta}_{\ell} \!\!&= \mathbb{E}(\tilde{\theta}_\ell - \theta_\ell), \!\!\!\!&
			\sigma^{\theta}_{\ell} \!\!&= \sqrt{\mathbb{V}(\tilde{\theta}_\ell - \theta_\ell)}, \ell \in \tilde{\mathcal{L}},		\end{aligned}
	\end{equation}
	where $\mathbb{E}(\cdot)$ and $\mathbb{V}(\cdot)$ denote the expectation and variance of estimated parameters ${\tilde{r}_{\ell}}$ and ${\tilde{\theta}_{\ell}}$, respectively. 
	Let $r_{\ell}^{\rm lb}$, $r_{\ell}^{\rm ub}$, $\theta_{\ell}^{\rm lb}$, and $\theta_{\ell}^{\rm ub}$ denote the lower and upper bounds of the search region for range and angle of path $\ell$, respectively, which are defined as follows
	\begin{equation}
	\label{eq:search_region}
	\begin{aligned}
		&r_{\ell}^{\rm lb} = 
		\max(\tilde{r}_\ell - \mu^{r}_{\ell} - 3\sigma^{r}_{\ell},Z_{\rm Fres}),\ell \in \tilde{\mathcal{L}},\\
		&r_{\ell}^{\rm ub} =
		\min(\tilde{r}_\ell - \mu^{r}_{\ell} + 3\sigma^{r}_{\ell},Z_{\rm Rayl}),\ell \in \tilde{\mathcal{L}},\\
		&\theta_{\ell}^{\rm lb} =
		\max(\tilde{\theta}_\ell - \mu^{\theta}_{\ell} - 3\sigma^{\theta}_{\ell},-1),\ell \in \tilde{\mathcal{L}},\\
		&\theta_{\ell}^{\rm ub} =
		\min(\tilde{\theta}_\ell - \mu^{\theta}_{\ell} + 3\sigma^{\theta}_{\ell},1), \ell \in \tilde{\mathcal{L}},
	\end{aligned}
	\end{equation}
	where $Z_{\rm Fres} = 0.5\sqrt{D^3/\lambda} $ and $Z_{\rm Rayl} = 2D^2/\lambda$ denote the Fresnel and Rayleigh distances, serving as the lower and upper bounds of the near-field region, respectively, with $D = (N-1)d$ representing the array aperture.
	Then the search region of the angle and range of each path can be obtained by the $3\sigma$ rule~\cite{10827219}, which defines the search region that contains the ground-truth parameters with high probability as follows
	\begin{equation} \label{B_search}
	\begin{aligned}
	\!\!\!\!\mathcal{B}\!\!=\!\!\big\{\!(\theta_{\ell},r_{\ell})|\theta_{\ell}^{\rm lb}\le\theta_{\ell}\le\theta_{\ell}^{\rm ub},
	r_{\ell}^{\rm lb}\le r_{\ell}\le r_{\ell}^{\rm ub},
	\forall \ell \in \tilde{\mathcal{L}}\big\}.\!\!\!\!\!
	\end{aligned}
	\end{equation}

	Note that we shall show in Section~\ref{sec:results} that the 3$\sigma$-criterion can effectively confine the search region around the coarse estimate $\tilde{\boldsymbol{\eta}}$, while ensuring that the ground-truth parameters lie within this region with high probability.

\section{Stage 2 for PSO-based Parameter Refinement} \label{subsec:pso_refinement} \vspace{-2pt}
For Stage 2, based on the coarse parameter estimation $\tilde{\boldsymbol{\eta}}$ in Stage 1, we propose a customized PSO method to refine the multi-path channel parameters for achieving improved estimation accuracy. 
Note that different from conventional PSO methods that randomly initialize particles over a full search space, our proposed method initializes particles around the coarse estimate $\tilde{\boldsymbol{\eta}}$ bounded by the confined search region $\mathcal{B}$ in Stage 1 (i.e.,~\eqref{B_search}), hence significantly reducing the search space and enhancing algorithm convergence speed.

\subsection{Problem Formulation}\vspace{-2pt}
To refine the multi-path channel parameters, we formulate an optimization problem to minimize the discrepancy between the actual received power pattern $\mathbf{p}_{\rm DFT}$ and the reconstructed power pattern based on the near-field multi-path channel model parameterized by $\boldsymbol{\eta}$ and $\mathbf{g}$. 
Specifically, let $\mathbf{B}({\boldsymbol{\eta}}) \in \mathbb{C}^{N \times (\tilde{L}+1)}$ define the reconstructed steering matrix based on the near-field multi-path channel model, which is given by
\begin{align}
		\!\!\!\mathbf{B}({\boldsymbol{\eta}}) \triangleq
		\big[{\mathbf{b}}({\theta}_0,{r}_0), {\mathbf{b}}({\theta}_1,{r}_1),
		\ldots,
		{\mathbf{b}}({\theta}_{\tilde{L}},{r}_{\tilde{L}})
		\big].\!\!
	\end{align}
Thus, the near-field multi-path channel in~\eqref{eq:total_channel_kappa} can be re-expressed as a function of the position parameters $\boldsymbol{\eta}$ and channel gains $\mathbf{g}$, i.e., $\mathbf{h}_{\rm near} = \mathbf{B}({\boldsymbol{\eta}})\mathbf{g}$.
Let $ \mathbf{\Phi}({\boldsymbol{\eta}}) \triangleq \mathbf{V}_{\rm DFT}^H\mathbf{B}({\boldsymbol{\eta}})$ denote the effective steering matrix. Then, given the confined search region $\mathcal{B}$ around the coarse estimate $\tilde{\boldsymbol{\eta}}$ in Stage 1, the problem of Stage 2 can be formulated as follows
\begin{align} \vspace{-2pt}
\textbf{(P4)}:
\min_{\boldsymbol{\eta},\mathbf{g}} \quad & \big\| \mathbf{p}_{\rm DFT} - |\mathbf{\Phi}(\boldsymbol{\eta}) \mathbf{g}|^2 \big\|_{2}^{2} \nonumber \\
\text{s.t.}\quad & \boldsymbol{\eta} \in \mathcal{B}. \label{eq:P1_cons} \vspace{-2pt}
\end{align}

	Note that Problem (\textbf{P4}) is a non-convex optimization problem and thus is generally challenging to solve, since 1) the modulus operation in the objective function, and 2) the non-linear coupling between $\boldsymbol{\eta}$ and $\mathbf{g}$ in the objective function.
	Moreover, the search dimension of Problem (\textbf{P4}) is $4(\tilde{L} + 1)$, hence making it computationally prohibitive to solve by conventional optimization methods when $\tilde{L}$ is large.
	To tackle these difficulties, we propose an inner-outer iterative optimization method to solve Problem (\textbf{P4}) efficiently, where the inner layer estimates the channel gains $\mathbf{g}$ given position parameters $\boldsymbol{\eta}$, while the outer layer optimizes the position parameters $\boldsymbol{\eta}$ using a customized PSO method.

	\underline{\textbf{Inner problem:}} Given any fixed position parameters $\boldsymbol{\eta}$, the inner layer problem of channel-gain optimization is given by 
	\begin{align}\label{eq:GS_problem}
		\textbf{(P5)}:~
		\min_{\mathbf{g}}~
		\big\|
		\mathbf{p}_{\rm DFT}-|\mathbf{\Phi}({\boldsymbol{\eta}})\mathbf{g}|^2
		\big\|_2^2. 
	\end{align}
	We denote $\bar{\mathbf{g}}$ as the optimized solution to Problem (P5) and define $G(\boldsymbol{\eta}) \triangleq \big\|
		\mathbf{p}_{\rm DFT}-|\mathbf{\Phi}({\boldsymbol{\eta}})\bar{\mathbf{g}}|^2$, which is the optimized objective function of Problem (P5) given any feasible $\boldsymbol{\eta}$.

	\underline{\textbf{Outer problem:}} 
	Based on the inner problem, the outer layer problem of position parameters optimization can be expressed as
	\begin{align} \vspace{-2pt}
	 	\textbf{(P6)}:~
	 	\min_{\boldsymbol{\eta}}~  
	 	G({\boldsymbol{\eta}})  \quad  
	 \text{s.t.}~\eqref{eq:P1_cons}. \nonumber  \vspace{-2pt}
	 \end{align}

	\vspace{-4pt}
	\subsection{Inner-layer for Channel Gain Estimation} \vspace{-2pt}
	Note that given any fixed position parameters $\boldsymbol{\eta}$, Problem (\textbf{P5}) is still a non-convex optimization problem, which is difficult to solve, since the objective function only depends on the received power, while the phase information is missing.
	To solve this phase retrieval problem, we propose a GS-based channel gain estimation method~\cite{10845209}, which is an alternating projection method that iteratively projects the  channel gain onto two domains until convergence, namely, a signal power domain with $\mathbf{p}_{\rm DFT}$ and a steering matrix domain with $\mathbf{\Phi}({\boldsymbol{\eta}})$,~respectively.

	Specifically, the GS algorithm iteratively refines the estimate of $\mathbf{g}$ by alternating between two steps based on a properly designed initialization.
	\begin{itemize}
		\item \textbf{Projection onto signal power domain.} 
		In the $i$-th iteration, given the current estimate of $\mathbf{g}$, denoted by $\mathbf{g}^{(i)}$, we first obtain the noiseless estimated signal as $\tilde{\mathbf{y}}^{(i)} = \mathbf{\Phi}({\boldsymbol{\eta}})\mathbf{g}^{(i)}$. 
		Then, we project $\tilde{\mathbf{y}}^{(i)}$ onto the signal power domain to enforce the power constraint by replacing the power of $\tilde{\mathbf{y}}^{(i)}$ with the received power $\mathbf{p}_{\rm DFT}$ while retaining the phase of $\tilde{\mathbf{y}}^{(i)}$.
		This results in a refined estimated signal $\hat{\mathbf{y}}^{(i)}$, which can be expressed as
		\begin{align}\label{project_y}
			\hat{\mathbf{y}}^{(i)} = \sqrt{\mathbf{p}_{\rm DFT}} \odot \frac{\tilde{\mathbf{y}}^{(i)}}{|\tilde{\mathbf{y}}^{(i)}|},
		\end{align}
		with $\odot$ denoting the element-wise multiplication.
		\item \textbf{Projection onto steering matrix domain.} Next, we project the estimated signal $\hat{\mathbf{y}}^{(i)}$ back onto the steering matrix domain to update the channel gain $\mathbf{g}^{(i+1)}$.
		This is achieved by solving the following least-squares  problem
		\begin{align}\label{project_LS}
			\mathbf{g}^{(i+1)} = \arg\min_{\mathbf{g}} \big\| \hat{\mathbf{y}}^{(i)} - \mathbf{\Phi}({\boldsymbol{\eta}})\mathbf{g} \big\|_2^2,
		\end{align}
		which admits a closed-form solution
		\begin{align}\label{project_g}
			\mathbf{g}^{(i+1)} = \mathbf{\Phi}^{\dagger}({\boldsymbol{\eta}}) \hat{\mathbf{y}}^{(i)},
		\end{align}
		with $\mathbf{\Phi}^{\dagger}({\boldsymbol{\eta}}) = \big(\mathbf{\Phi}^{H}({\boldsymbol{\eta}})\mathbf{\Phi}({\boldsymbol{\eta}})\big)^{-1}\mathbf{\Phi}^{H}({\boldsymbol{\eta}})$ denoting the pseudo-inverse of $\mathbf{\Phi}({\boldsymbol{\eta}})$.
	\end{itemize}

	\vspace{-4pt}
	\begin{remark}[Initialization of GS algorithm]\emph{
		The initialization of $\mathbf{g}$ in the GS algorithm can significantly affect its convergence and estimation accuracy, due to the non-convex nature of Problem (\textbf{P5}).
	To address this issue, we design a dedicated spectral initialization method~\cite{7130654} to provide a rigorous and high-quality initial estimate of $\mathbf{g}$ by exploiting the statistical structure of the received power pattern.
	Specifically, let $\boldsymbol{\varphi}^T_{n}$ denote the $n$-th row of the effective steering matrix $\mathbf{\Phi}({\boldsymbol{\eta}})$, such that $\mathbf{\Phi}({\boldsymbol{\eta}}) = [\boldsymbol{\varphi}^T_{1}, \boldsymbol{\varphi}^T_{2}, \ldots, \boldsymbol{\varphi}^T_{N}]^T$. We first construct a positive semi-definite correlation matrix, $\mathbf{M}$, as a weighted sum of outer products of the effective steering vectors, where the weights correspond to the received power pattern $\mathbf{p}_{\rm DFT} = [p_1, \ldots, p_N]^T$, which is given by
	$
		\mathbf{M} = \frac{1}{N} \sum_{n=1}^{N} {p}_{n} \boldsymbol{\varphi}_{n} \boldsymbol{\varphi}_{n}^H.
	$
	According to the principle of spectral initialization, the principal eigenvector of $\mathbf{M}$ aligns with the direction of the true channel vector. To obtain this direction, we perform the eigenvalue decomposition (EVD) of $\mathbf{M}$ to obtain its eigenvalues and eigenvectors as
	$
		\mathbf{M} = \mathbf{E} \boldsymbol{\Lambda} \mathbf{E}^H,
	$	
	where $\mathbf{E} = [\mathbf{e}_0, \mathbf{e}_1, \ldots, \mathbf{e}_{\tilde{L}}]$ contains the eigenvectors, and $\boldsymbol{\Lambda} = \mathrm{diag}(\lambda_0, \lambda_1, \ldots, \lambda_{\tilde{L}})$ contains the eigenvalues sorted in a descending order, i.e., $\lambda_0 \geq \lambda_1 \geq \ldots \geq \lambda_{\tilde{L}}$.
	The eigenvector $\mathbf{e}_0$, corresponding to the largest eigenvalue $\lambda_0$, captures the spatial structure (relative phase and amplitude) of the channel gain. 
	However, as $\mathbf{e}_0$ is unit-norm, i.e., $\|\mathbf{e}_0\|_2 = 1$, it lacks the absolute scale information.
	To recover the absolute strength, we perform a power scaling that minimizes the~discrepancy between the received power pattern and the reconstructed one, which is given by
	$\beta = \sqrt{ \frac{\sum_{n=1}^{N} p_n}{\sum_{n=1}^{N} |\boldsymbol{\varphi}^{H}_n\mathbf{e}_0|^2} }$.
	Finally, the initialization for $\mathbf{g}^{(0)}$ of the GS algorithm is set as
	$
		\mathbf{g}^{(0)} = \beta \mathbf{e}_0.
	$
	}\end{remark} \vspace{-2pt}

	Given the above initialization for $\mathbf{g}^{(0)}$, which ensures linear convergence with high probability~\cite{7130654}, the GS algorithm proceeds to refine the estimate of $\mathbf{g}$ via alternating projections in~\eqref{project_y} and \eqref{project_g}.
	This process continues until convergence is achieved, yielding the estimated channel gains $\bar{\mathbf{g}} $.

\vspace{-10pt}
\subsection{Outer-layer for Position Parameters Estimation}\vspace{-2pt}
	The outer Problem \textbf{(P6)} is generally challenging to solve due to the non-convexity of the effective steering matrix $\mathbf{\Phi}({\boldsymbol{\eta}})$ w.r.t. $\boldsymbol{\eta}$.
	Traditional gradient-based methods may easily get stuck in local optima within highly complex search spaces.
	To address this issue, we employ a gradient-free heuristic-based PSO algorithm to obtain a high-quality solution to Problem \textbf{(P6)}, which has been shown to achieve exceptional global search capabilities and robustness in tackling non-convex estimation problems~\cite{11329408}.

Specifically, for the proposed PSO-based algorithm, we first initialize $M_{\rm PSO}$ particles. 
Let $\mathcal{P}_{\text{PSO}}^{(0)} \triangleq \{\boldsymbol{\eta}_1^{(0)}, \dots, \boldsymbol{\eta}_{M_{\text{PSO}}}^{(0)}\}$ denote the candidate parameter set of the multi-path channel, where $\boldsymbol{\eta}_1^{(0)}$ is set as the coarse estimate from Stage 1, i.e., $\boldsymbol{\eta}_1^{(0)} = \tilde{\boldsymbol{\eta}}$, while the remaining particles $\{\boldsymbol{\eta}_m^{(0)}\}_{m=2}^{M_{\rm PSO}}$ are randomly initialized within the search region $\mathcal{B}$ to ensure diversity in the initial population.
Moreover, the associated velocities $\mathcal{U}_{\text{PSO}}^{(0)} \triangleq \{\mathbf{u}_1^{(0)}, \dots, \mathbf{u}_{M_{\text{PSO}}}^{(0)}\}$ denote the update vectors for each particle, which are initialized to zero, i.e., $\mathbf{u}_m^{(0)} = \mathbf{0}, \forall m \in \{1,\ldots,M_{\rm PSO}\}$.

Then, we define $f_{\text{fit}}(\cdot)$ as a \textit{fitness function}, which guides the optimization process by evaluating how well the power pattern  constructed by the candidate parameter vector $\boldsymbol{\eta}$ matches the received power pattern.
Mathematically, $f_{\text{fit}}(\cdot)$ is defined as 
\begin{align} \label{fitness_function}
f_{\text{fit}}(\boldsymbol{\eta}) 
= 
\big\|
		\mathbf{p}_{\rm DFT}-|\mathbf{\Phi}({\boldsymbol{\eta}})\bar{\mathbf{g}}|^2
		\big\|_2^2
+ 
\zeta J(\boldsymbol{\eta}),
\end{align}
where $\zeta>0$ controls the penalty weight and $J(\boldsymbol{\eta})$ is a boundary penalty designed to penalize any candidate parameters that violate the bounds of the search region $\mathcal{B}$, given by
\begin{align}
J(\boldsymbol{\eta})
&= 
\sum_{\ell=0}^{\tilde{L}}\Big[ f_{\rm p}(r_{\ell}, r_\ell^{\rm ub}, r_\ell^{\rm lb})+ f_{\rm p}(\theta_\ell, \theta_\ell^{\rm ub}, \theta_\ell^{\rm lb})\Big].
\end{align}
Herein, $f_{\rm p}(x, x^{\rm ub}, x^{\rm lb})$ denotes the normalized quadratic penalty function, defined as
\begin{align}
f_{\rm p}(x, x^{\rm ub}, x^{\rm lb})
=
\frac{(x - x^{\rm ub})_+^2 + (x^{\rm lb} - x)_+^2}
{(x^{\rm ub} - x^{\rm lb})^2},
\end{align}
where $(y)_+ \triangleq \max\{0,y\}$ is the positive operation, ensuring that penalties are incurred only when the bounds are violated.

Then, for each particle in the population, the velocity and candidate parameters are updated based on both its individual best and the global best positions.
Specifically, the individual best position $\boldsymbol{\eta}_{m,\text{pbest}}^{(t)}$ and the global best position $\boldsymbol{\eta}_{\text{gbest}}^{(t)}$ at iteration $t$ are determined as
\begin{equation} \label{PSO_best}
\begin{aligned}
\boldsymbol{\eta}_{m,\text{pbest}}^{(t)} 
&= 
\arg\!\!\min_{\!\!\!\!\!\tau \in \{0,\ldots,t\}} 
f_{\text{fit}}\big(\boldsymbol{\eta}_m^{(\tau)}\big), \forall m \in \{1,\ldots,M_{\text{PSO}}\},\\
\boldsymbol{\eta}_{\text{gbest}}^{(t)} 
~~~&= 
\arg\!\!\!\!\!\!\!\min_{\!\!\!\!\!m \in \{1,\ldots,M_{\text{PSO}}\}}
f_{\text{fit}}\big(\boldsymbol{\eta}_{m,\text{pbest}}^{(t)} \big).
\end{aligned}
\end{equation}
Thus, the velocity and position of the $m$-th particle at iteration $t$ are updated as follows
\begin{equation} \label{PSO_updated}
	\begin{aligned}
	\mathbf{u}_m^{(t+1)} 
	&= 
	\rho \,\mathbf{u}_m^{(t)}
	+ \varrho_1\tau_1\big(\boldsymbol{\eta}_{m,\text{pbest}}^{(t)} - \boldsymbol{\eta}_m^{(t)}\big) \\
&\quad\quad\qquad
+ \varrho_2\tau_2\big(\boldsymbol{\eta}_{\text{gbest}}^{(t)} - \boldsymbol{\eta}_m^{(t)}\big),\\
\boldsymbol{\eta}_m^{(t+1)} 
&= 
\boldsymbol{\eta}_m^{(t)} + \mathbf{u}_m^{(t+1)},
\end{aligned}
\end{equation}
where $\rho$ denotes the inertia weight, while $\varrho_1$ and $\varrho_2$ represent the learning factors for the cognitive and social components, respectively. The terms $\tau_1,\tau_2 \sim \mathcal{U}[0,1]$ are random coefficients introducing stochastic exploration.
 
Following the above procedures, after $T_{\rm PSO}$ iterations, a high-quality solution to Problem \textbf{(P4)} is obtained, which is denoted as $ \bar{\boldsymbol{\eta}}= \boldsymbol{\eta}_{\rm{gbest}}^{(T_{\rm PSO})} $.

\begin{remark}[Algorithm convergence and computational complexity]\label{prop:complexity}\rm
First, for the proposed hybrid learning-and-optimization method, its convergence is determined by the parameter refinement in Stage 2. 
Since the objective function of Problem \textbf{(P4)} is non-increasing over PSO iterations and is lower-bounded by zero, the convergence of the proposed algorithm is guaranteed.

Next, consider the computational complexity of proposed method.
For the U-Net model inference in Stage 1, its computation complexity is proportional to the size of the network (i.e., the number of layers, the feature sizes, and the spatial dimensions of each layer).
Thus, the computation complexity of Stage 1 is in the order of $C_{\text{DL}} = \mathcal{O}(\sum_{i=1}^{5} D_i D_{i-1} N/2^{i-1}K)$, where $K$ is the kernel size of the convolutional layer in DoubleConv block.
For typical U-Net-based architectures, the inference time of $C_{\text{DL}}$ is generally in the order of microseconds~\cite{10478627,10682562}, which is negligible compared to iterative optimizations. 
For Stage 2, the computational complexity of PSO refinement 
is dominated by the GS iterations in the inner-layer for channel gain estimation and the evaluation of the fitness function in \eqref{fitness_function} in the outer-layer for each particle.
Specifically, the computational complexity of the projection in~\eqref{project_y} is in the order of $\mathcal{O}(N)$, which is dominated by the element-wise multiplication between $\mathbf{p}_{\rm DFT}$ and $\tilde{\mathbf{y}}^{(i)}$.
Meanwhile, the computational complexity of the projection in~\eqref{project_LS} is in the order of $\mathcal{O}(N \tilde{L})$, which is dominated by the matrix multiplication between $\mathbf{\Phi}^{\dagger}({\boldsymbol{\eta}})$ and $\hat{\mathbf{y}}^{(i)}$.
As such, the computational complexity of the GS algorithm is in the order of $C_{\text{GS}} = \mathcal{O}\big(I_{GS}(N + N\tilde{L})\big)$, where $I_{GS}$ is the required number of iterations of the GS algorithm for convergence.
The outer-layer PSO optimization requires evaluating the fitness function for each particle in each iteration, which has a computational complexity of $ \mathcal{O}(N\tilde{L})$ due to the matrix multiplication between $\mathbf{\Phi}({\boldsymbol{\eta}})$ and $\bar{\mathbf{g}}$ in \eqref{fitness_function}.
Thus, the computational complexity of the PSO refinement is in the order of $C_{\text{PSO}} = \mathcal{O}\big(M_{\text{PSO}} T_{\text{PSO}} (I_{GS}(N + N\tilde{L}) + N\tilde{L})\big)$.
Based on the above, the total computational complexity of the proposed hybrid learning-and-optimization beam training algorithm can be expressed as $C_{\text{Total}} =  \mathcal{O}\big(M_{\text{PSO}} T_{\text{PSO}} N (I_{GS}(1 + \tilde{L}) + \tilde{L})\big)$.

Note that the coarse estimation in Stage~1 significantly narrows the feasible search region, making it possible to use a much smaller number of particles and iterations to achieve satisfactory convergence compared to global space search.
\end{remark}

\vspace{-5pt}
\section{Numerical Results} \label{sec:results} \vspace{-2pt}
In this section, we present numerical results to demonstrate the effectiveness of our proposed hybrid learning-and-optimization beam training method.

\vspace{-5pt}
\subsection{System Setup and Benchmark Schemes}\vspace{-2pt}
The system parameters are set as follows unless otherwise specified.  
We consider an XL-array with $N = 256$ antennas operating at a frequency of $30$~GHz, thus the wavelength is $\lambda = 0.01$~m.
The channel is generated based on the near-field multi-path channel model in \eqref{eq:total_channel_kappa}, where the number of NLoS paths is uniformly set as $L \in \{2,3,4\}$.
The spatial angles $\{\theta_\ell\}$ and ranges $\{r_\ell\}$ are independently and uniformly distributed within $\theta_\ell \sim \mathcal{U}(-0.5 ,0.5) $ and $r_\ell \sim \mathcal{U}(8,38)$ m.  
The complex-valued channel gain $g_0$ and $g_{\ell}$ are modeled as $g_0 = \sqrt{\frac{\kappa}{\kappa+1}}\frac{\lambda}{4\pi r_0} e^{-\jmath \frac{2\pi}{\lambda} r_0}$ and $g_\ell \sim \mathcal{CN}(0,\sigma^2_l) , \ell \in \mathcal{L}$, where $\sigma_l = \frac{1}{\sqrt{L({\kappa+1})}}\frac{\lambda}{4\pi r_{0}}$ and $\kappa$ denotes the Rician factor varying from 0 dB to 30 dB~\cite{10500334,10123941}.
The reference received signal-to-noise ratio (SNR) is defined as $\text{SNR} = {P_{\rm t}||\mathbf{h}_{\rm near}||^2}/{\sigma^2}$, where the transmit power $P_{\rm t} = 10$ dBm  and the noise power $\sigma^2 = -80$ dBm, respectively.
The size of dataset is 500000, among which 80\% is used for training, 10\% for validation, and 10\% for testing.
All evaluations are conducted using Python 3.9 and PyTorch 2.2 on a computer equipped with an Intel Xeon Platinum 8336C CPU @ 2.30GHz and a graphics processing unit (GPU) (NVIDIA RTX 4090 with 24GB memory).
Other parameters are summarized in~Table~\ref{tab:hyperparameters}.

\begin{table}[t]
\centering
\caption{Simulation hyperparameter configuration}\vspace{-2pt}
\label{tab:hyperparameters}
\begin{tabular}{|c|c|c|}
\hline
\textbf{Stage} & \textbf{Hyperparameter} & \textbf{Value} \\
\hline
\multirow{7}{*}{Stage 1} & 
Feature dimensions, $\{D_i\}_{i=1}^5$ & [64, 128, 256, 512, 1024] \\\cline{2-3} 
&Weighting coefficients, $\{\alpha_i\}_{i=1}^3$ & [1, 1, 1] \\\cline{2-3} 
&Initial learning rate & $10^{-3}$ \\ \cline{2-3} 
&Batch size & 256 \\ \cline{2-3} 
&Number of epochs & 1000 \\ \cline{2-3} 
&Convolution kernel size & 3 \\ \cline{2-3} 
&Optimizer & Adam \\ 
\hline
\multirow{6}{*}{Stage 2} &
Number of particles, $M_{\text{PSO}}$ & 50 \\ \cline{2-3} 
&Number of iterations, $T_{\text{PSO}}$ & 100 \\ \cline{2-3} 
&Inertia weight, $\rho$ & 0.7 \\ \cline{2-3} 
&Individual learning factor, $\varrho_1$ & 1.5 \\ \cline{2-3} 
&Global learning factor, $\varrho_2$ & 1.5 \\ \cline{2-3} 
&Penalty weight, $\zeta$ & 100 \\
\hline 
\end{tabular}
\vspace{-16pt}
\end{table}

For performance comparison, we consider the following benchmark schemes:
\begin{itemize}
	\item \textbf{Perfect-CSI-based beamforming}: The beamforming vector is designed based on perfect CSI, which serves as an upper bound for performance comparison.
	\item \textbf{LoS-based scheme}: This scheme assumes that only a strong LoS path exists and infers channel parameters based on the received power pattern, which first estimates user angle based on a DFT codebook, and then estimates user range with a polar-domain codebook~\cite{9913211}.
	\item \textbf{CNN-based scheme}: This scheme adopts the conventional CNN-based near-field beam training method proposed in~\cite{9903646}, where the CNN is trained to directly map the received power pattern to the polar-domain codebook.
	\item \textbf{Far-field beam training}: This scheme performs beam training based on the DFT codebook by selecting the beams with the largest received powers, where the number of selected beams is set to the number of paths assumed to be known \textit{a priori}.
\end{itemize}

Besides, we also consider the following variants of the proposed method to evaluate the effectiveness of each stage:
\begin{itemize}
	\item \textbf{Coarse estimation of Stage 1}: The coarse estimates $\tilde{\theta}_\ell, \tilde{r}_\ell$ obtained from Stage 1 are used to reconstruct the near-field multi-path channel for beamforming designs, where channel gains are estimated via the GS method.
	\item \textbf{PSO with full space}: For this scheme, we consider the PSO-based parameter estimation, where the full parameter space is searched without prior information.
	\item \textbf{Proposed method ($1\sigma$ boundary)}: This scheme applies the proposed method with a search boundary defined by the 1$\sigma$-criterion in \eqref{eq:search_region}, which is more restrictive than the 3$\sigma$-criterion used in the proposed method.
\end{itemize}

\vspace{-10pt}
\subsection{Convergence Performance Analysis}
\begin{figure*}[t]	\vspace{-8pt}
	\centering
	\hspace{-16pt}
	\subfigure[Train loss and test loss for three components of the loss function in~\eqref{eq:total_loss_fun} versus epochs.]{	\label{fig:Loss_vs_epochs}
	\includegraphics[width=0.3\textwidth]{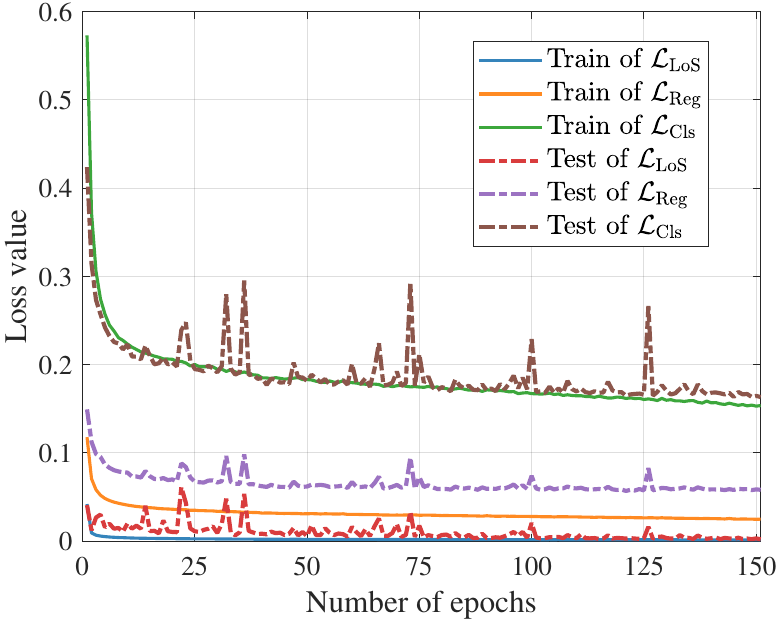}} \hspace{0pt}
	\subfigure[Fitness function value of~\eqref{fitness_function} versus iterations.]{	\label{fig:PSO_convergence}
	\includegraphics[width=0.298\textwidth]{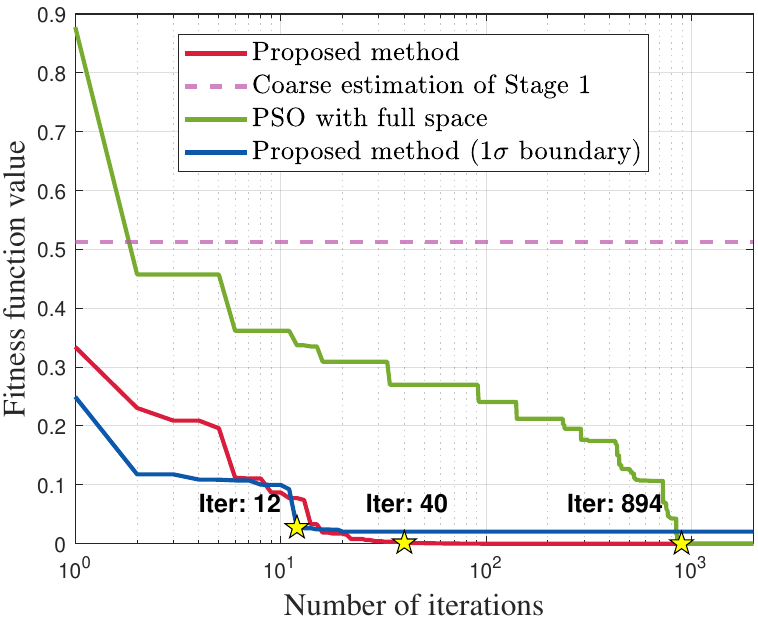}}\hspace{3pt}
	\subfigure[Average convergence runtime versus SNR.]{	\label{fig:runtime_vs_snr}
	\includegraphics[width=0.3\textwidth]{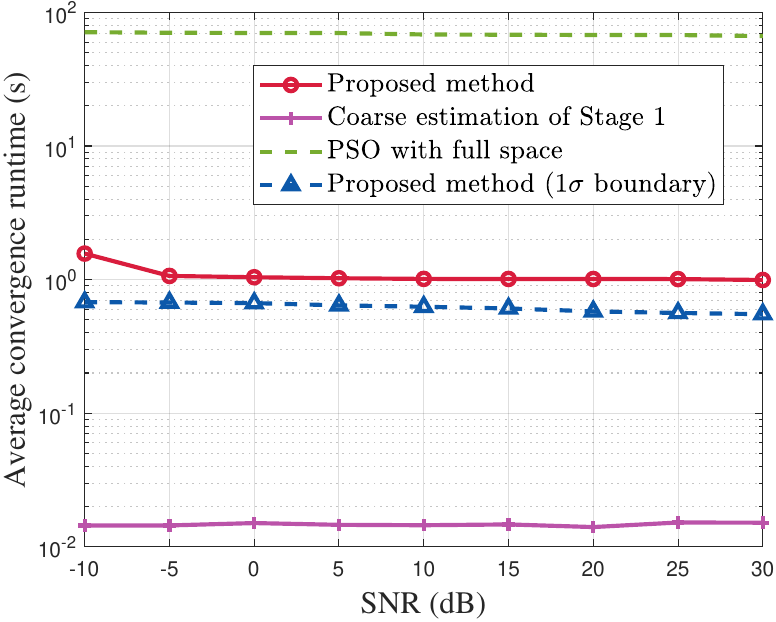}}\hspace{-6pt}
	\vspace{-8pt}
	\caption{\centering Convergence behavior and runtime analysis of the proposed method.}
	\label{fig:convergence}\vspace{-16pt}
\end{figure*}

\vspace{-2pt}
Fig.~\ref{fig:Loss_vs_epochs} illustrates the convergence behavior of the training and testing losses for the proposed U-Net model across epochs. 
Specifically, it depicts three distinct loss components defined in~\eqref{eq:total_loss_fun}. 
It is observed that all loss curves exhibit a sharp descent during the initial training phase before stabilizing, indicating that the model effectively learns the features from the complex near-field multi-path power pattern. 
Furthermore, the testing loss curves are close to the training loss curves with minimal divergence, demonstrating strong generalization capabilities and robustness against overfitting.

Fig.~\ref{fig:PSO_convergence} illustrates the convergence performance of the proposed PSO-based parameter refinement algorithm by plotting the fitness function value against the number of iterations. 
It is observed that the proposed method initiates with a significantly lower fitness value compared to the conventional PSO method with full space. 
This advantage results from the high-quality coarse estimates from Stage 1, enabling the construction of a confined search region for effective particle initialization.
Consequently, the proposed method exhibits superior convergence characteristics, reaching a stationary point rapidly after 40 iterations. 
In contrast, the conventional PSO method with full space exhibits a significantly slower convergence rate, necessitating a substantial increase in the number of iterations to approach a satisfactory solution. 
On the other hand, while the proposed method with $1\sigma$-criterion boundary exhibits the fastest convergence, it suffers at a suboptimal fitness value. This indicates that an overly restrictive search boundary may fail to cover the true parameter range, thereby preventing the algorithm from locating the global optimum.

Fig.~\ref{fig:runtime_vs_snr} compares the average convergence runtime of the different beam training schemes versus SNR, which is measured until convergence.
It is observed that the PSO with full space scheme incurs the longest runtime to convergence (around $10^{2}$s) among all methods, which is due to its exhaustive search over the entire high-dimensional parameter space without any prior information to guide the optimization process.
In contrast, the coarse estimation exhibits the shortest runtime (around  $10^{-2}$s), as its complexity is limited to a single forward inference pass of the U-Net. 
Crucially, the proposed method maintains a low runtime (around  $10^{0}$s) that is significantly shorter than the conventional PSO benchmark method. 
This efficiency stems from the learning-based coarse estimation, which effectively confines the search region and thereby allowing the optimization-based parameter refinement to converge rapidly with significantly fewer iterations.

\begin{figure*}[t]	
	\centering
	\hspace{-7pt}
	\subfigure[Estimation results for position parameters.]{	\label{fig:parameter_estimation_accuracy}
	\includegraphics[width=0.3\textwidth]{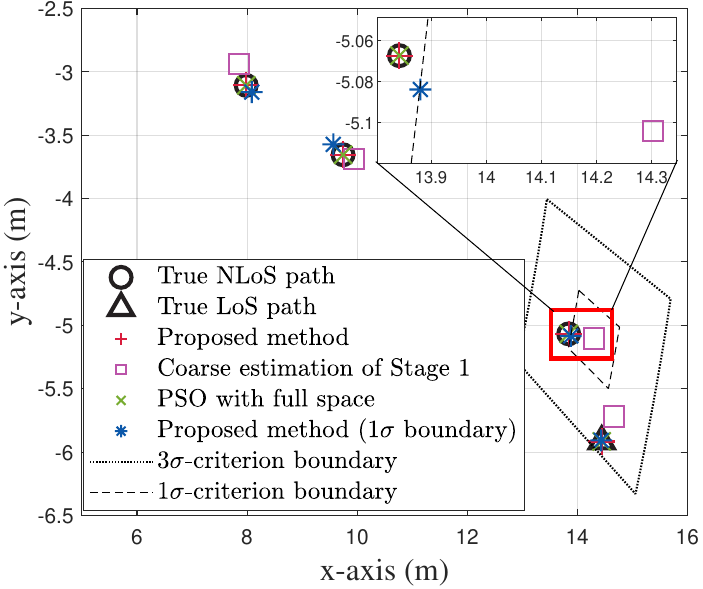}}\hspace{-4pt}
	\subfigure[Comparison of reconstructed power patterns.]{	\label{fig:Beam_pattern_comparison}
	\includegraphics[width=0.325\textwidth]{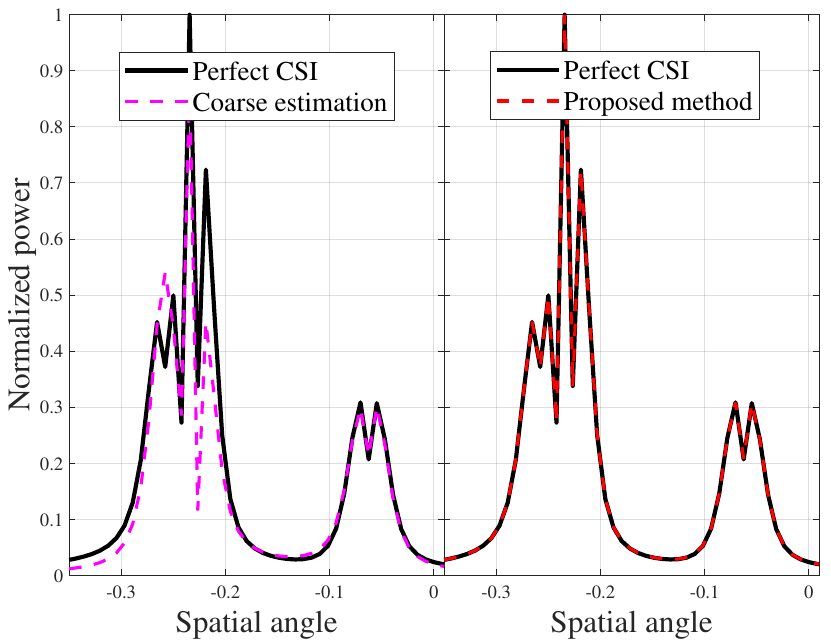}}\hspace{-4pt}
	\subfigure[Path detection accuracy versus SNR.]{	\label{fig:path_detection_vs_snr}
	\includegraphics[width=0.31\textwidth]{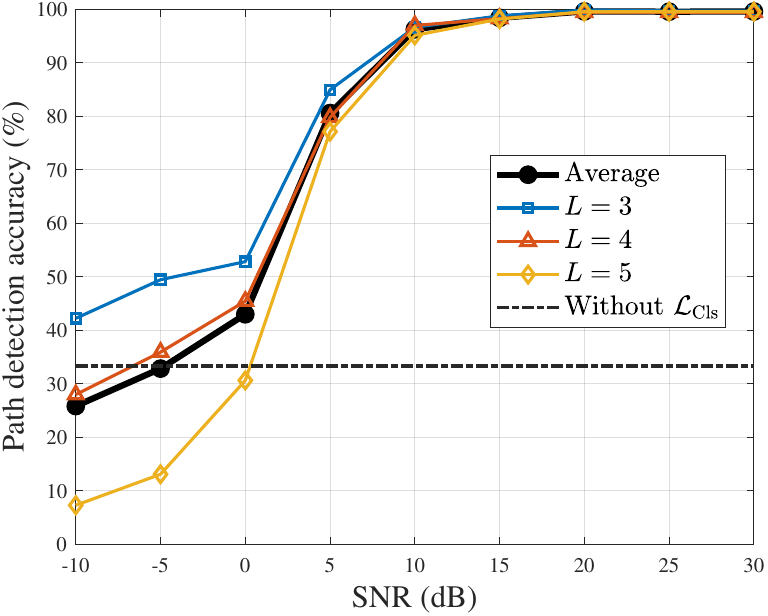}} 
	\vspace{-8pt}
	\caption{\centering Key parameters estimation accuracy, reconstructed power pattern comparison, and path detection accuracy versus SNR.} \vspace{-14pt}
\end{figure*}

\vspace{-10pt}
\subsection{Parameter Estimation Comparison} \vspace{-2pt}
Fig.~\ref{fig:parameter_estimation_accuracy} demonstrates the parameter estimation accuracy by plotting the estimated coordinates of LoS and NLoS paths in the Cartesian system. 
Herein, the LoS path and NLoS paths are spatially close in the angular domain, typically inducing severe signal superposition, making it highly challenging to distinguish individual components based on the received power pattern alone.
It is observed that the proposed coarse estimation method successfully disentangles these paths and obtains coarse estimates of targets, which demonstrates superior capability of the proposed U-Net model in handling the strong ambiguity caused by signal superposition.
Based on this, the subsequent PSO-based refinement stage of the proposed method achieves superior localization precision, with the refined estimates closely aligning with the true positions. 
Furthermore, a closer observation reveals the effectiveness of the 3$\sigma$-criterion for search space construction: the 3$\sigma$-criterion boundary successfully encompasses the true parameters, ensuring that the global optimum is included within the confined search region for the subsequent optimization stage. 
In contrast, the overly restrictive search region defined by the  1$\sigma$-criterion excludes the true path location, which makes the PSO algorithm converge to a biased boundary estimate, thereby failing to accurately recover the channel parameters.

Fig.~\ref{fig:Beam_pattern_comparison} compares the normalized power patterns reconstructed by different estimation schemes against the ideal power pattern obtained from perfect CSI. 
Specifically, the normalized power pattern reconstructed by coarse estimation successfully captures the profile of power pattern, indicating that the U-Net model effectively resolves the number and locations of dominant paths.
However, the reconstructed multi-path amplitude exhibits noticeable deviations, which are due to residual errors in the neural network output.
In contrast, the power pattern reconstructed by the proposed method exhibits an exceptional alignment with that under perfect CSI, accurately estimating the multi-path components.

Fig.~\ref{fig:path_detection_vs_snr} evaluates the accuracy of the proposed method in estimating the number of paths versus SNR. 
It is observed that the estimation accuracy improves significantly as the SNR increases. 
Specifically, in the low-SNR regime (e.g., -10 dB), the detection of multiple NLoS paths is challenging due to the strong noise interference, resulting in lower accuracy, particularly for scenarios with a larger number of paths (e.g., 5 paths). 
However, as the SNR improves, the estimation accuracy improves rapidly and approaches nearly 100\% when the SNR exceeds 10 dB. 
Besides, without the classification loss in~\eqref{Loss_Cls}, the model will always estimate the maximum number of paths, which results in a constant accuracy as the frequency of the maximum number of paths in the dataset (i.e., around 30\%). 
This result validates the effectiveness of the proposed path existence estimation loss function in~\eqref{Loss_Cls}.

\begin{figure*}[t]	
	\vspace{-10pt}
	\centering\hspace{-4pt}
	\subfigure[NMSE versus SNR.]{	\label{fig:nmse_comp}
\includegraphics[width=0.292\textwidth]{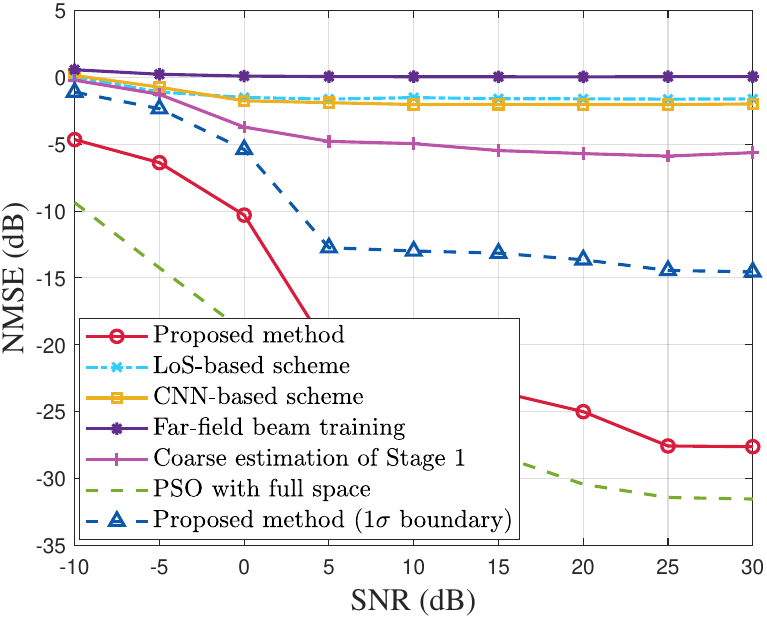}} \hspace{-3pt}
	\subfigure[Angle estimation RMSE versus SNR.]{	\label{fig:rmse_theta_comp}
		\includegraphics[width=0.3\textwidth]{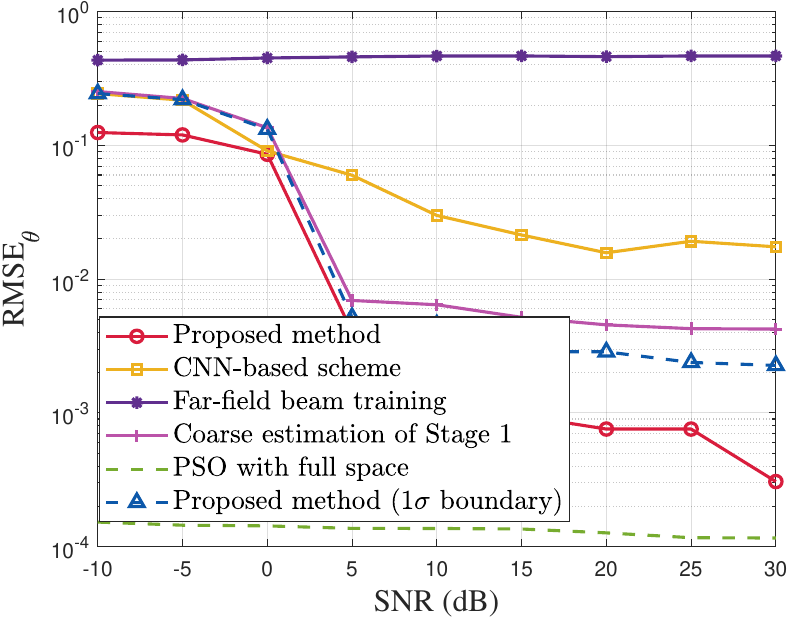}} \hspace{-2pt}
	\subfigure[Range estimation RMSE versus SNR.]{	\label{fig:rmse_range_comp}
		\includegraphics[width=0.3\textwidth]{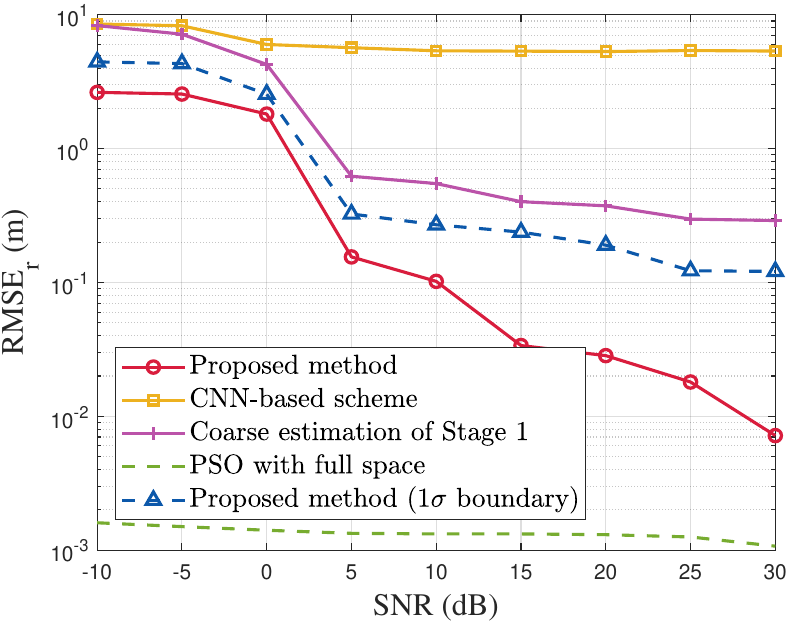}}\hspace{4pt}
		\vspace{-8pt}
	\caption{\centering Estimation errors of channel, angle and range versus SNR.}
	\label{fig:error_vsSNR}\vspace{-18pt}
\end{figure*}

\vspace{-10pt}
\subsection{Estimation Error Comparison}
\begin{figure*}[t]	
	\centering
	\subfigure[Achievable rate versus SNR.]{	\label{fig:rate_comp}
	\includegraphics[width=0.3\textwidth]{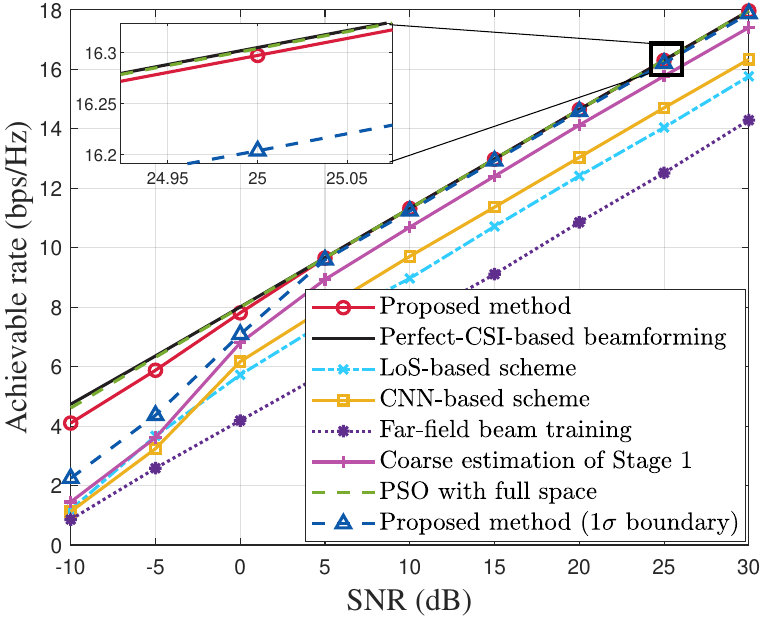}} 
	\subfigure[Achievable rate versus number of antennas.]{	\label{fig:rate_comp_antennas}
	\includegraphics[width=0.3\textwidth]{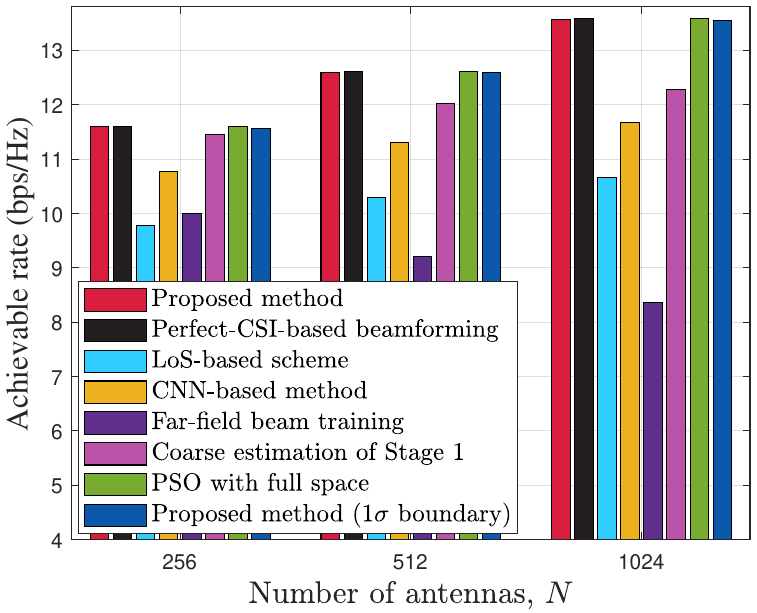}} 
	\subfigure[Achievable rate versus SNR under the 3GPP TR 38.901 UMa scenario.]{	\label{fig:rate_comp_3gpp}
	\includegraphics[width=0.3\textwidth]{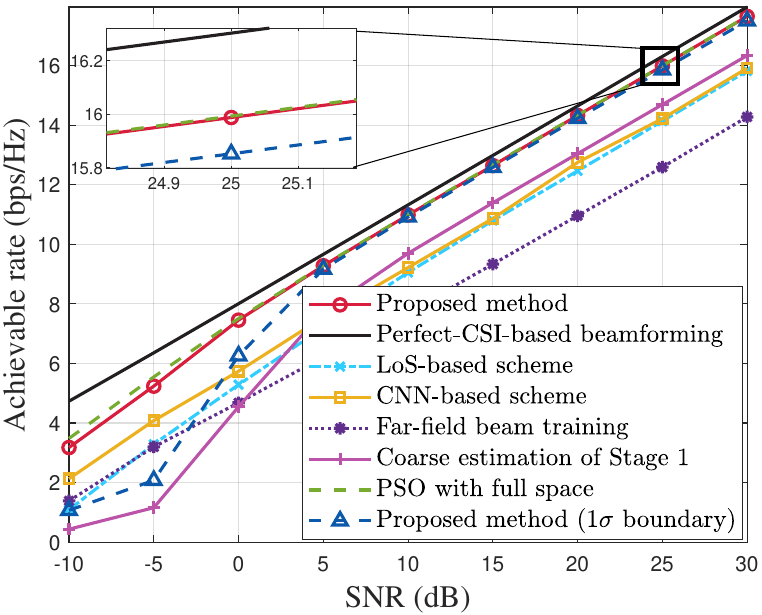}} 
	\vspace{-8pt}
	\caption{\centering Achievable rate versus SNR and number of antennas} \vspace{-17pt}
\end{figure*}

\vspace{-2pt}
Let the normalized MSE (NMSE) be defined as $\text{NMSE} = \mathbb{E}[\|\mathbf{h}_{\rm near} - \bar{\mathbf{h}}_{\rm near}\|^2/\|\mathbf{h}_{\rm near}\|^2]$, with $\bar{\mathbf{h}}_{\rm near}$ denoting the reconstructed  channel based on the estimated parameters.
Fig.~\ref{fig:nmse_comp} evaluates the normalized MSE (NMSE) of the reconstructed channel versus SNR for the different schemes.
It is observed that the coarse estimation of Stage 1 and the CNN-based scheme suffer from a pronounced error floor, saturating at a relatively high NMSE (e.g., around -5 dB and -2 dB).
The saturation is due to the inherent resolution limitations of classification-based models, which prevent them from achieving high-precision continuous parameter recovery.
In contrast, the NMSE of the proposed method decreases with the increase of SNR, reaching a minimum of approximately $-27$ dB in the high-SNR regime.
Noting that while the PSO with full space scheme exhibits slightly better estimation precision than the proposed method, this difference can be negligible in terms of rate when SNR is higher than 5 dB (to be shown in Fig.~\ref{fig:rate_comp}).

Figs.~\ref{fig:rmse_theta_comp} and \ref{fig:rmse_range_comp} compare the root MSE (RMSE) of angle and range estimation versus SNR for different beam training schemes, respectively, where the estimation RMSEs of angle and range are defined as $\text{RMSE}_{\theta} = \sqrt{\mathbb{E}[\|\boldsymbol{\theta} - \bar{\boldsymbol{\theta}}\|^2]}$ and $\text{RMSE}_{r} = \sqrt{\mathbb{E}[\|\mathbf{r} - \bar{\mathbf{r}}\|^2]}$, respectively.
It is observed that both the coarse estimation of Stage 1 and the CNN-based scheme saturate at a noticeable error floor even in high-SNR conditions.
In contrast, the RMSE for both angle and range of the proposed method decreases monotonically with increasing SNR, effectively overcoming the error floor observed in pure learning-based benchmarks. 
This validates the effectiveness of the optimization-based refinement  in correcting the residual errors from the coarse estimation stage.
Moreover, the proposed method with $1\sigma$-criterion boundary saturates at a suboptimal level compared to that with the $3\sigma$-criterion boundary. 
This indicates that an overly restricted boundary may exclude the true channel parameters, thereby preventing the optimization algorithm from converging to the global~optimum.

\vspace{-10pt}
\subsection{Achievable Rate Comparison} \vspace{-2pt}
Fig.~\ref{fig:rate_comp} evaluates the achievable rate given the estimated channel versus SNR for different beam training schemes. 
First, the benchmark schemes neglecting NLoS components or spherical wavefronts (e.g., the LoS-based scheme and Far-field beam training) suffer severe performance degradation due to significant channel model mismatch.
Second, compared to the CNN-based scheme, the coarse estimation of Stage 1 achieves a substantial performance gain across the entire SNR regime.
This improvement is attributed to the superior feature extraction capabilities of the U-Net architecture, where the encoder-decoder structure with skip connections allows for effective fusion of multi-scale features, thereby preserving the fine-grained spatial details of the multi-path channel that are often lost in the conventional CNN-based scheme.
Third, the proposed method with $1\sigma$ demonstrates a noticeable performance improvement over the coarse estimation of Stage 1 but eventually saturates at suboptimal. 
This indicates that the search boundary defined by the $1\sigma$ criterion is overly restrictive, which may fail to include the true parameters, thereby preventing the optimization algorithm from reaching the global optimum.
Finally, the PSO with full space scheme achieves an achievable rate that tightly approaches the perfect-CSI-based beamforming upper bound, where the method runs a significantly large number of iterations to ensure convergence.
In comparison, the proposed method similarly achieves this near-optimal performance with low complexity.

Besides, Fig.~\ref{fig:rate_comp_antennas} investigates the effect of the number of antennas on the achievable rate.
The path ranges are fixed at $70$ m, which ensures that the user and scatterers remain within the near-field region as the array aperture expands.
As expected, the achievable rate for all near-field schemes improves with the number of antennas, benefiting from the increased array gain and spatial resolution.
In contrast, the far-field beam training method suffers from performance degradation as the antenna number increases due to the model mismatch caused by the more pronounced near-field effect.
Moreover, one can observe that our proposed method outperforms other methods across different numbers of antennas, maintaining a negligible performance gap relative to the upper bound.

To further validate the effectiveness of the proposed scheme under more practical propagation conditions, we additionally provide the achievable rate comparison under a 3GPP-based simulation setup~\cite{10078317}. Specifically, QuaDRiGa is employed to generate the channel dataset according to the 3GPP TR 38.901 UMa scenario at 24 GHz~\cite{11327450,6758357,3GPP38901v1920}. 
The corresponding simulation results are shown in Fig.~\ref{fig:rate_comp_3gpp}.
It is observed that the proposed method still achieves rate performance close to its upper bound based on perfect CSI, significantly outperforming benchmark schemes.
Even under the low-SNR regime (e.g., -10 dB), the proposed hybrid method 
still achieves a noticeable performance gain over benchmark schemes, which demonstrates its robustness in practical near-field complex scattering environments.
Compared with the simulation results in Fig.~\ref{fig:rate_comp}, the performance gap between the proposed method and its  upper bound based on perfect CSI is slightly larger under the 3GPP-based simulation setup, which is due to the discrepancy between the ideal near-field channel model and the complex practical channel modelling.

\vspace{-6pt}
\section{Conclusion} \label{sec:conclusion}
In this paper, we studied near-field beam training under multi-path channel conditions. 
In order to overcome the limitations of conventional DFT-based methods, we proposed a two-stage hybrid learning-and-optimization method for near-field multi-path beam training. 
In particular, a U-Net model was employed in the first stage to provide a coarse estimation of the key multi-path parameters, while a PSO-based optimization method was applied in the second stage to achieve high-precision refinement within a confined search region. 
Lastly, numerical results verified the effectiveness of the proposed hybrid method in significantly improving estimation accuracy compared with existing schemes.

\vspace{-6pt}
	\bibliographystyle{IEEEtran}
	\bibliography{Ref_title.bib}
\end{document}